\documentclass[aps,pra,floatfix,superscriptaddress, showpacs, nofootinbib]{revtex4}

\usepackage{graphicx}
\usepackage{graphics}
\usepackage{amssymb}
\usepackage{amsmath}
\usepackage{epsfig}
\usepackage{latexsym}
\usepackage{color}
\usepackage{rotating}
\usepackage{subfigure}

\def\ketc[#1]{\vert #1 \rangle}
\def\brac[#1]{\langle #1 \vert}

\def\gbl{\bar{\gamma}_L}
\def\gbr{\bar{\gamma}_R}
\newcommand{\expect}[1]{\langle{#1}\rangle}

\newcommand{\beq}{\begin{equation}}
\newcommand{\eeq}{\end{equation}}
\newcommand{\bqa}{\begin{eqnarray}}
\newcommand{\eqa}{\end{eqnarray}}

\newcommand{\fuller}{C$_{60}$}
\begin{document}
\title{Spin-detection in a quantum electromechanical shuttle system}
\author{J. Twamley}
\email{jtwamley@ics.mq.edu.au} \affiliation{Centre for Quantum Computer Technology,
Macquarie University, Sydney, New South Wales 2109, Australia}

\author{D. W. Utami}
\affiliation{School of Physical Sciences, University of Queensland,
St Lucia, QLD 4072, Australia}

\author{H. -S. Goan}
\affiliation{Department of Physics, National Taiwan University,
Taipei 106, Taiwan, ROC}

\author{G. Milburn}
\affiliation{Centre for Quantum Computer Technology, The University
of Queensland, St Lucia, QLD 4072, Australia}

\begin{abstract}
We study the electrical transport of a harmonically-bound,
single-molecule \fuller shuttle operating in the Coulomb blockade
regime, i.e. single electron shuttling. In particular we examine the
dependance of the tunnel current on an ultra-small stationary force
exerted on the shuttle. As an example we consider the force exerted
on an endohedral N@\fuller by the magnetic field gradient generated by a
nearby nanomagnet. We derive a Hamiltonian for the full shuttle
system which includes the metallic contacts, the spatially dependent
tunnel couplings to the shuttle, the electronic and motional degrees
of freedom of the shuttle itself and a coupling of the shuttle's
motion to a phonon bath. 
We analyse the resulting quantum master equation and find that, 
due to the exponential dependance of the
tunnel probability on the shuttle-contact separation, 
the current  is highly sensitive to very small forces. In particular we predict that 
the spin state of the endohedral electrons of N@\fuller  in a large magnetic
gradient field can be distinguished from the resulting current signals within a few tens of nanoseconds. 
This effect could prove useful for the detection of
the endohedral spin-state of individual paramagnetic molecules  such
as N@\fuller  and P@\fuller, or the detection of very small static forces acting on a \fuller  shuttle.\\
\end{abstract}

\maketitle

\section{Introduction}
The determination of the spin-state of a single electron spin has
attracted much attention recently \cite{Nature}. Following the
ground breaking experiments of Park {\it et al.}, in the observation
of quantized motional transitions in the conductance of a
single-molecule \fuller transistor \cite{mceuen}, interest in
quantum electromechanical systems, or QEMS,  has blossomed
\cite{electromechanical}. In this work we wish to examine whether a
QEMS, with the inclusion of spin degrees of freedom, or a Spin-QEMS,
can be used to detect the spin-state of the included spin. We have
in mind the particular case of Nitrogen or Phosphorous endohedrally
doped C$_{60}$, N@\fuller or P@C$_{60}$.  This material has
previously been shown to possess unpaired endohedrally trapped
electrons, in a quartet ground state, $S=3/2$, and which display
exceedingly long transverse relaxation (or $T_2$), times. This
material has been considered for use in quantum information
processing devices but may have uses in other areas related to
classical spintronic memories.

As a first exploration one can consider the original experiment of
Park {\it et al.} using the paramagnetic endohedral, like P@\fuller,
instead of C$_{60}$. It is known that the electronic spin degrees of
freedom reside in the endohedral electrons which are concentrated
isotropically at the centre of the \fuller cage in P@\fuller
\cite{twamley}. In the experiment of \cite{mceuen}, the \fuller
oscillates in a Harmonic potential at THz frequencies. For the
Spin-QEMS system of a P@\fuller molecule as a movable island in a
single molecule transistor, one must be capable of engineering a
coupling between the state of the endohedral spin and the
conductance properties of the transistor. This can be accomplished
through the imposition of a large magnetic field gradient $\partial
B/\partial x$, along the plane of the transistor. The interaction
between the endohedral spin and this gradient field will impart a
small force on the molecule which will in turn, cause a slight shift
in the equilibrium position within the damped Harmonic oscillator
trapping potential, shifting the molecule towards one of the
contacts. This small shift will lead to a change in conductance
properties. It will be shown that if the configuration is
asymmetric, i.e. the mobile island is closer to one lead than the
other,  the resulting current depends {\em exponentially}, on this
small spatial displacement.  In the case of  a tight trapping,  observed
in Park {\it et al.}, \cite{mceuen}, the
resulting spin-dependent change in conductance is
extremely small and unobservable. If we instead configure for the largest
achievable magnetic field gradient (obtained for
instance in MRFM devices), by placing a nanoscopic permanent magnet
near to the mobile island, we also find that due to the large difference between the 
energy scales of the endohedral electronic spin flip (GHz),
and the island's harmonic motion in \cite{mceuen}, (THz),  
the island's spin-dependent position shift is minute and is unobservable in the current.

\begin{figure}[h!]
\includegraphics[scale=0.8]{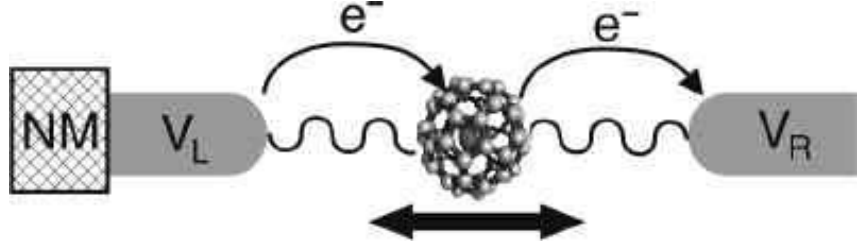}
\caption{Schematic of the N@\fuller molecular shuttle. $NM$ labels the
nanomagnet, while the left and right tunnel contacts are held a the
voltages $V_L$ and $V_R$ respectively. } \label{schematic}
\end{figure}

Guided by these observations we instead consider the more ``floppy''
Spin-QEMS of a shuttle device where the restoring oscillator now
operates in the GHz frequency regime (see Figure \ref{schematic}). MRFM cantilevers fabricated
out of single Si crystals have recently been generated which have
resonant frequencies in the GHz range \cite{cantilevers}. The
resulting device could be similar to that outlined by Isacsson
\cite{Isacsson}. It may also be possible to engineer non-conducting
flexible linker molecules which will connect the shuttle molecule to
the metallic contacts and have resonant frequencies in the GHz
range.

We thus consider the system of a movable shuttle or island, which is
harmonically bound to oscillate between two metallic contacts in the
presence of significant motional damping. We further assume that we
operate in the Coulomb blockade region where only one electron can
reside on the shuttle at one time. For the particular case of an
endohedral N\{P\}@\fuller island, as in \cite{mang1}, we make the
assumption that this injected electron distributes itself
isotropically around the surface of the \fuller cage thus averaging
out any magnetic dipole coupling between the spin state of the
injected electron and the spin state of the endohedral electrons. We
also assume that the  injected electrons are not spin polarised and
the resulting force experienced by the shuttle due to the interaction
between the spin of the injected electrons and the magnetic gradient
will average out while the force on the shuttle due to the
endohedral electrons interacting with the magnetic field remains
static.  We therefore neglect the random spin-force contribution of
the injected electrons in what follows.

%We now estimate the force that will be experienced by the shuttle due to the interaction of the endohedral spin and a magnetic field gradient created by the placing of a small nanoscopic ferromagnetic particle nearby.
Before going into the model in detail, we can understand, in a
simple but surprisingly robust manner the underlying reasons why we
expect to achieve an {\em exponential} dependence of the current on
the island's displacement from its equilibrium position. The current
through a simple source-fixed island-drain, two-stage sequential
tunneling process \cite{buttiker}, in terms of the left and right
tunnel rates $\gamma_L$, $\gamma_R$, is given by \beq
i\sim\frac{\gamma_L\gamma_R}{\gamma_R+\gamma_L}\;\;. \label{i} \eeq
If we now take into account the exponential dependence of these
tunnel rates on the position of the (now mobile) island, and letting
$\delta$ (positive) be the small displacements of the island from
its equilibrium position to the right (to the drain), we can set
\beq \gamma_L\sim \gamma_L^0e^{-\lambda
\delta},\;\;\;\gamma_R\sim\gamma_R^0e^{+\lambda \delta}\;\;, \eeq
and thus the right tunnel rate, from the island to the drain,
increases due to the smaller separation while the left tunnel rate
decreases due to the larger separation. This is reversed if the motion
is to the left ($\delta$ negative), or towards the source:
\beq
\gamma_L\sim \gamma_L^0e^{+\lambda |\delta|},\;\;\;\gamma_R\sim\gamma_R^0e^{-\lambda |\delta|}\;\;.
\eeq
Inserting these into (\ref{i}), we easily get:
\beq
\Omega\equiv
\frac{i^{positive\;\;displacement}}{i^{negative\;\;displacement}}=\frac{\gamma_L^0e^{-\lambda
|\delta|}+\gamma_R^0e^{+\lambda |\delta|}}{\gamma_L^0e^{+\lambda
|\delta|}+\gamma_R^0e^{-\lambda |\delta|}}=\frac{1+F e^{+2\lambda
|\delta|}}{e^{+2\lambda |\delta|}+F}\;\;,\label{ratio} \eeq where
$F=\gamma_R^0/\gamma_L^0$.  If $F=1$ then $\Omega=1$, and there is
no difference between the currents in the two cases. When 
%$F \gg 1$,
$ F \gg e^{+2\lambda |\delta|}$,
 $\Omega \sim e^{+2\lambda |\delta|}$, while if 
 %$F  \ll 1$,
$F \ll e^{- 2\lambda |\delta|}$,
  $\Omega
\sim e^{-2\lambda |\delta|}$,
and in either of these asymmetric cases the two currents can be very different from each other if $\lambda |\delta|$ is appreciable.\\
  \\
In the sections below, we analyse the classical and quantum models
of this system. Following the above simple rationale, we can find a
realistic parameter range where the right-hand tunnel current shows
a large (exponential), dependance on small spin-dependent island
position displacements. This is confirmed below using  low-resolution numerical simulations of
both the ``semiclassical'' and quantum systems, and through
high-resolution quantum simulations. Besides demonstrating the
  two different current ``signals''  corresponding to the spin force, we
must also determine the noise present within the system to estimate
the signal to noise ratio or so-called Fano factor. Following
standard methods we derive the current spectral noise density and
from this the signal to noise properties of the coupled system.
Using all this we finally deduce the measurement time required to
distinguish the spin state from the current signals and can find
device parameters where this acquisition time will be several tens
of nanoseconds.

\section{Magnetic Field Gradient}
\begin{figure}[h!]
\includegraphics[scale=0.5]{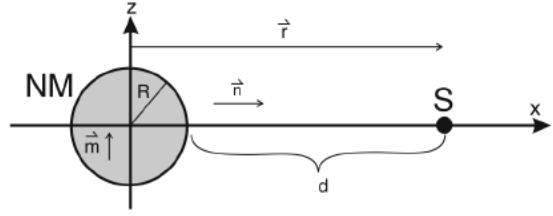}
\caption{Schematic of the nanomagnet, a small spherical
ferromagnetic particle $NM$, of radius $R$, a distance $d$ from the
shuttle $S$.} \label{magnet}
\end{figure}

We now estimate the magnitude of the maximum magnetic field gradient
we can achieve at the location of the shuttle. We follow a similar
derivation in Magnetic Resonance Force Microscopy (MRFM)
\cite{Bermen}. The Hamiltonian for the shuttle's endohedral spin,
$\vec{S}$, interacting with a magnetic field which varies along the
$x-$axis (see Figure \ref{magnet}), $\vec{B}(x)$, is
$H=-\vec{\mu}\cdot\vec{B}=-\mu_Bg_eB(x)S_z$, while the resulting
force on the shuttle is given by $F_{spin}=\mu_Bg_eS_zdB(x)/dx$. If
we consider placing a nanoscopic ferromagnetic spherical particle of
radius $R$ a distance $d$ from the shuttle, one has \beq
\vec{B}_{Ferro}=\frac{\mu_0}{4\pi}\frac{3\vec{n}(\vec{m}\cdot
\vec{n})-\vec{m}}{r^3}\;\;,\label{Bferro} \eeq where
$\mu_0=4\pi\times 10^{-7}$H/m, $\vec{m}$ is the magnetic moment of
the ferromagnetic particle, $\vec{r}$ is the vector connecting the
particle to the shuttle and $r=|\vec{r}|=R+d$. For a sphere the
magnitude of the magnetic moment is $|\vec{m}|=4/3\pi R^3M$, where
$M$ is the magnetization of the ferromagnetic material. Following
this we have $\vec{B}_{Ferro}=-\mu_0R^3M/r^3 \hat{z}$. As the
shuttle moves along the $x-$axis, the separations $d$ and $r$,
between the shuttle and the magnet change. Setting the distance
between the shuttle and the magnet's centre to be $r\equiv
R+d_0+x=\bar{x}+x$, where $x$ is the small oscillation of the
shuttle, $d_0$ is the equilibrium separation, and $\bar{x}=R+d_0$,
one can derive \beq F_{spin}=\frac{3\mu_Bg_e\mu_0M
S_z}{\bar{x}}\left(\frac{R}{\bar{x}}\right)^3\;\;.\label{Fspin} \eeq
For an iron ferromagnet, $\mu_0M=2.2$Tesla, and choosing $d_0=5$nm,
$R=5$nm, $\bar{x}=10$nm, $S_z=\pm 3/2$ (for P@\fuller), one obtains
$F_{spin}\sim \pm 1\times 10^{-15}$N.

Using this we can re-examine the spin Hamiltonian and expand to
first order in $x$ about the equilibrium shuttle position,
$H=-\mu_Bg_eS_zB_z(r)=-\mu_Bg_eS_zB_z(\bar{x}+x)\sim
H_0-F_{spin}x+{\cal O}(x^2)$, where $H_0$ is a constant which we
drop. From (\ref{Fspin}), above we set \beq
H_{spin}=-\tilde{\chi}S_zx\;\;,\label{Hspin} \eeq where \beq
\tilde{\chi}\equiv \frac{3\mu_eg_e\mu_0
M}{\bar{x}}\left(\frac{R}{\bar{x}}\right)^3\;\;.\label{chidef1} \eeq

The Hamiltonian for the Coloumb-energy of the charged island/shuttle
can be written as \beq H_{shuttle\;electronic}=\hbar \omega_I
c^\dagger c\;\;, \eeq where $c$ represents the operator which
annihilates an electron on the shuttle, and $c^\dagger c$ is the
number of injected electrons on the shuttle, which in the blockade
situation is either 0 or 1.

\section{Electrostatic Force}
In most studies of electron shuttles \cite{electromechanical}, one
considers the shuttle to be driven primarily by the electrostatic
force exerted on the shuttle through the interaction of the shuttle
charge and an electric field present between the source and drain
metallic contacts created by a potential difference $V=V_L-V_R$,
across the inter-contact gap $D$. This gives an electric field of
strength $E=V/D$.

%In standard treatment of electron shuttling
%\cite{electromechanical}, the shuttle moves towards the left contact
%(source), and acquires negative change though electron tunneling,
%then moves rapidly to the right contact (drain), under the
%electrostatic force generated by this negative charge in the
%electric field of the contacts. At the drain the shuttle looses its
%negative charge but then acquires an excess of holes or a net
%positive charge, and the shuttle, again under the electrostatic
%force, moves back to the source where the process repeats.

In a standard treatment of electron shuttling \cite{utami}, the
shuttle acquires the electron at the average displacement of zero
slightly towards the source and continues to move towards the
source. It then undergoes electrostatic force to be propelled back
towards the drain and loses the electron at a slightly displaced
position towards the drain while continuing to move towards it.
Again under electrostatic force the shuttle moves back towards the
source and repeats the process. In this case no harmonic restoring
force is required to sustain the shuttle instability. However in our
case, in the situation where our shuttle is operating in the Coulomb
blockade regime, the shuttle can acquire a single excess negative
charge near the source and thus be forced electrostatically to the
drain, but once the shuttle dumps its electron at a displaced
position near the drain there is no electrostatic force which can
bring this shuttle back to the source contact to repeat the shuttle
sequence. Thus in our case a Harmonic restoring force is required to
sustain the shuttling action.

In more detail, $F_{ES}=qE=-eE$, where $E=-dV/dx\sim (V_R-V_L)/D=
-V/D$, if $V_L=-V_R=V/2$. Setting the potential due to this
electrostatic force to be $\phi_{ES}(x)$, $F_{ES}=-d\phi_{ES}/dx$,
and $\phi_{ES}(x)=-\int F_{ES}dx=-qEx=|e|Ex=-|e|Vx/D$. Setting
$\tilde{\eta}\equiv V|e|/D$, we have \beq
H_{ES}=-\tilde{\eta}\,x\;\;. \label{HES} \eeq

\section{Harmonic Oscillator}
The Hamiltonian of the Harmonic oscillator is dealt with in the
standard manner. Setting the annihilation operator for motional
quanta to be, $\sqrt{2}a=x/x_0+ip/p_0$, where $x_0=\sqrt{\hbar/(m
\omega_0)}$ and $p_0=\sqrt{m\omega_0\hbar}$ are the extents in $x$
and $p$, of the  ground state wave function of the harmonic
oscillator, we have \beq
H_{HO}=\frac{p^2}{2m}+\frac{k_0}{2}x^2=\frac{p^2}{2m}+\frac{m\omega_0^2}{2}x^2=\hbar\omega_0\,a^\dagger
a\;\;.\label{HHO} \eeq

\section{Complete Hamiltonian}
We are now in a position to write out the complete Hamiltonian for
the drain, source and motional, spin and electronic degrees of
freedom of the shuttle:
\begin{eqnarray}
H&=&\hbar\omega_I\,c^\dagger c\label{H1}\\
&-&\chi\,S_z(a^\dagger+a)/\sqrt{2}\label{H2}\\
&-&\eta(a^\dagger+a)c^\dagger c/\sqrt{2}\label{H3}\\
&+&\hbar \omega_0 a^\dagger a\label{H4}\\
 &+&\hbar\sum_{k_l}\,\omega^L_{k_l}\,a^{\dagger\,L}_{k_l}a^L_{k_l}\label{H5}\\
 &+&\hbar\sum_{k_r}\,\omega^R_{k_r}\,a^{\dagger\,R}_{k_r}a^R_{k_r}\label{H6}\\
 &+&\hbar\sum_{k_b}\,\omega_{k_b}\,b^{\dagger}_{k_b}b_{k_b}+(\sum_{k_i}\,a^\dagger b_{k_i}+{\rm h.c.})\label{H7}\\
 &+&\sum_{k_{tl}}\,(T^L_{k_{tl}}\hat{E}_L(\hat{x})a^L_{k_{tl}}\,c^\dagger+{\rm h.c.} )\label{H8}\\
 &+&\sum_{k_{tr}}\,(T^R_{k_{tr}}\hat{E}_R(\hat{x})a^R_{k_{tr}}\, c^\dagger+{\rm h.c.} )\label{H9}
 \end{eqnarray}
   where
  $ \chi\equiv\tilde{\chi}x_0$, and $\eta\equiv \tilde{\eta}x_0$. In the above: (\ref{H1}) is the Coloumb self-energy of the shuttle, (\ref{H2}) is the spin-dependent potential energy, (\ref{H3}) is the electrostatic potential energy, (\ref{H4}) is the Harmonic oscillator self-energy, (\ref{H5}) and (\ref{H6}) are the self-energies of the Fermi baths in the source and drain contacts, (\ref{H7}) is the self-energy and coupling of a phonon bath to the motion of the shuttle, (\ref{H8}) is the tunneling interaction between the source contact and the shuttle, while (\ref{H9}) is the tunneling interaction between the shuttle and the drain contact.

  We note the operators $\hat{E}_{L/R}(\hat{x})$, in the tunneling terms. The amplitude for tunneling depends exponentially on the spatial separation between the source/drain and the shuttle. One can assume the forms
  \beq
  \hat{E}_L(\hat{x})=\gamma_L^{1/2}e^{-\tilde{\lambda}_L\,x}\;\;,\qquad
  \hat{E}_R(\hat{x})=\gamma_R^{1/2}e^{+\tilde{\lambda}_R\,x}\;\;, \label{myexponential}
  \eeq
  where $x$ is the displacement of the shuttle from its equilibrium position.
In the following we assume the source and drain are identical in
shape and material and thus $\tilde{\lambda}_L=
\tilde{\lambda}_R=\tilde{\lambda}$, where for Gold contacts,
$\tilde{\lambda}^{-1}\sim 3$ Angstroms. As above we set
$\lambda\equiv\tilde{\lambda}x_0$ to get \beq
\hat{E}_{L/R}=\gamma_{L/R}^{1/2}\,\exp(\mp \lambda (a^\dagger
+a)/\sqrt{2}) \;\;. \eeq

\section{Displacement Picture}
As we mentioned above, the small static spin force exerted on the
shuttle, in the absence of the electrostatic force, will cause a
slight shift in the shuttle's equilibrium position within the
Harmonic trapping potential. We now move to a frame of reference
where we shift this displacement back to the origin. This can be
done by considering the transformation 
\beq \rho\rightarrow
\rho^\prime=D(\alpha)\rho D^\dagger(\alpha)\;\;,\label{displacement} 
\end{equation}
where $D(\alpha)\equiv \exp (\alpha a^\dagger - \alpha^* a)$, is the standard displacement operator for operators satisfying  $[a^\dagger,a]=1$.
By choosing $\alpha=-(\chi S_z)/(\sqrt{2}\hbar \omega_0)$,
which corresponds to the physical displacement of the shuttle's
equilibrium position due to the spin force $F_{spin}$, of \beq
\tilde{\delta}=-\frac{F_{spin}}{m\omega_0^2}\;\;, \eeq we can apply
the displacement transformation (\ref{displacement}), to the above
full Hamiltonian to get
\begin{eqnarray}
H^\prime&=&\hbar\omega_I\,c^\dagger c \pm \delta \eta c^\dagger c \label{HH1}\\
&-&\eta(a^\dagger+a)c^\dagger c/\sqrt{2}\label{HH3}\\
&+&\hbar \omega_0 a^\dagger a\label{HH4}\\
 &+&\hbar\sum_{k_l}\,\omega^L_{k_l}\,a^{\dagger\,L}_{k_l}a^L_{k_l}\label{HH5}\\
 &+&\hbar\sum_{k_r}\,\omega^R_{k_r}\,a^{\dagger\,R}_{k_r}a^R_{k_r}\label{HH6}\\
 &+&\hbar\sum_{k_b}\,\omega_{k_b}\,b^{\dagger}_{k_b}b_{k_b}+(\sum_{k_i}\,a^\dagger b_{k_i}+{\rm h.c.})\label{HH7}\\
 &+&\sum_{k_{tl}}\,(T^L_{k_{tl}}\hat{E}_L(\hat{x}^\prime)a^L_{k_{tl}}\,c^\dagger+{\rm h.c.} )\label{HH8}\\
 &+&\sum_{k_{tr}}\,(T^R_{k_{tr}}\hat{E}_R(\hat{x}^\prime)a^R_{k_{tr}}\, c^\dagger+{\rm h.c.} )\label{HH9}
 \end{eqnarray}
 where now
 \begin{eqnarray}
 \hat{E}_L(\hat{x}^\prime)=\gamma_L^{1/2}\,\exp\left(-\lambda [(a^\dagger+a)/\sqrt{2}\mp \delta]\right)\equiv \gamma_L^{1/2}E_L\;\;,\label{EL}\\
 \hat{E}_R(\hat{x}^\prime)=\gamma_R^{1/2}\,\exp\left(+\lambda [(a^\dagger+a)/\sqrt{2}\mp \delta]\right)\equiv\gamma_R^{1/2}E_R\;\;,\label{ER}
 \end{eqnarray}
 where $\delta\equiv \tilde{\delta}/x_0$.

 \section{Master Equation}
 Assuming that the time scales for the couplings to the source, drain and phonon baths are faster than those associated with the Harmonic oscillator and electrostatic potential, one can derive the following master equation \cite{utami}:
 \begin{eqnarray}
 \frac{d\rho}{dt}&=&-i\omega_0[a^\dagger a,\rho]+i\frac{\eta}{\hbar}[(a^\dagger+a)c^\dagger c/\sqrt{2},\rho]\\
 &+&\gamma_L\left\{f_L(\omega_I \mp \delta \eta){\cal D}[c^\dagger E_L]\rho+(1-f_L(\omega_I \mp \delta \eta)){\cal D}[cE_L]\rho\right\}\\
 &+&\gamma_R\left\{f_R(\omega_I \mp \delta \eta){\cal D}[c^\dagger E_R]\rho+(1-f_R(\omega_I \mp \delta \eta)){\cal D}[cE_R]\rho\right\}\\
&+&\kappa(\bar{N}+1){\cal D}[a]\rho +\kappa \bar{N}{\cal
D}[a^\dagger]\rho\;\;,
\end{eqnarray}
where the super-operator ${\cal D}$ is defined to be ${\cal
D}[A]B\equiv ABA^\dagger-\frac{1}{2}(A^\dagger A B+BA^\dagger A)$.
The mean excitation of the thermal bath is $\bar{N} = 1/\exp(\hbar
\omega_0/k_B T - 1)$  and $f(x)=(\exp (x/(k_BT))+1)^{-1}$, is the
Fermi factor where
  \begin{eqnarray}
  f_L(\omega_I \mp \delta \eta)=f(\omega_I \mp \delta \eta-\mu_L)=f(\omega_I \mp \delta \eta-V/2)\;\;,\\
  f_R(\omega_I \mp \delta \eta)=f(\omega_I \mp \delta \eta-\mu_R)=f(\omega_I \mp \delta \eta+V/2)\;\;
  \end{eqnarray}
with a dependence on the bias voltage (through the chemical
potential).  Notice the dependency of the Fermi factor on the spin
force. If the displacement factor $\delta$ is large enough, 
 the Fermi energies corresponding to the two spin directions will become significantly different  
and large enough to be observable at low temperatures. If one supposes that the bias voltage can be tuned to be
 between these two energies such that only one Fermi
factor (corresponding to spin up say), is near unity while the other is nearly vanishing, it should be
possible to differentiate the two spins based on the detection or absence of
current. However with such low bias,
we believe the noise in the system will be overwhelming and
we thus proceed by assuming that the system is in large bias such
that this effect is negligible.

  Now at milli-Kelvin temperatures, $k_BT\sim 0.2\mu$eV, while $\omega_I\sim 5$meV. So if $V>10$meV then $f_L\rightarrow 1$, while $f_R\rightarrow 0$, giving the simpler form for the Master equation:
  \begin{eqnarray}
 \frac{d\rho}{dt}&=&-i\omega_0[a^\dagger a,\rho]+i\frac{\eta}{\hbar}[(a^\dagger+a)c^\dagger c/\sqrt{2},\rho]\label{mymaster}\\  
 &+&\gamma_L{\cal D}[c^\dagger E_L]\rho +\gamma_R{\cal D}[cE_R]\rho \nonumber\\ 
&+&\kappa(\bar{N}+1){\cal D}[a]\rho +\kappa \bar{N}{\cal D}[a^\dagger]\rho\;\;.\nonumber
\end{eqnarray}

 \section{Expectation Values}
 From the master equation (\ref{mymaster}), we can derive the following differential equations:

 \begin{eqnarray}
 \frac{d\langle c^\dagger c\rangle }{dt}&=&\gamma_L\langle \hat{E}_L^\dagger E_L(1-c^\dagger c)\rangle -\gamma_R\langle E_R^\dagger \hat{E}_Rc^\dagger c\rangle\;\;,\\\label{dcdt}
 \frac{d\langle x/x_0\rangle}{dt}&=&\omega_0\langle\frac{p}{p_0}\rangle-\frac{\kappa}{2}\langle\frac{x}{x_0}\rangle\;\;,\\
 \frac{d\langle p/p_0\rangle}{dt}&=&-\omega_0\langle\frac{x}{x_0}\rangle+\bar{F}_{ES}\langle c^\dagger c\rangle -\frac{\kappa}{2}\langle \frac{p}{p_0}\rangle\;\;,
 \end{eqnarray}
 where
 \beq
 \bar{F}_{ES}\equiv \frac{V|e|}{Dp_0}=\frac{|F_{ES}|}{p_0}
 \eeq
 These equations can be further tidied by going to natural time units associated with the Harmonic oscillator, i.e. $\tau=\omega_0 t$, and by defining
 \beq
 \bar{\gamma}_{L/R}=\frac{\gamma_{L/R}}{\omega_0}\;\;, \qquad\bar{\kappa}=\frac{\kappa}{2\omega_0}\;\;,\qquad\bar{\chi}_{ES}=\frac{\bar{F}_{ES}}{\omega_0}\;\;,\label{rescaled}
 \eeq
 we get

 \begin{eqnarray}
 \frac{d\langle c^\dagger c\rangle }{d\tau}&=&\bar{\gamma}_Le^{\pm
 2\lambda\delta} \langle e^{-2\lambda x/x_0}(1-c^\dagger c)\rangle
 -\bar{\gamma}_R e^{\mp 2\lambda \delta} \langle e^{+2\lambda x/x_0}
 c^\dagger c\rangle\;\;,\\ \frac{d\langle
 x/x_0\rangle}{d\tau}&=&\langle\frac{p}{p_0}\rangle-\bar{\kappa}\langle\frac{x}{x_0}\rangle\;\;,\\
 \frac{d\langle
 p/p_0\rangle}{d\tau}&=&-\langle\frac{x}{x_0}\rangle+\bar{\chi}_{ES}\langle
 c^\dagger c\rangle -\bar{\kappa}\langle \frac{p}{p_0}\rangle\;\;.
 \end{eqnarray}

 The above equations cannot be solved as they are not closed. One can break the correlations in the semiclassical regime of high damping. When this is done we finally obtain:

 \begin{eqnarray}
 \frac{d\langle c^\dagger c\rangle }{d\tau}&=&\bar{\gamma}_Le^{\pm 2\lambda\delta} e^{-2\lambda \langle x/x_0 \rangle}(1-\langle c^\dagger c\rangle) -\bar{\gamma}_R e^{\mp 2\lambda \delta}  e^{+2\lambda \langle x/x_0 \rangle } \langle c^\dagger c\rangle\;\;,\label{averagec}\\
 \frac{d\langle x/x_0\rangle}{d\tau}&=&\langle\frac{p}{p_0}\rangle-\bar{\kappa}\langle\frac{x}{x_0}\rangle\;\;,\nonumber\\
 \frac{d\langle p/p_0\rangle}{d\tau}&=&-\langle\frac{x}{x_0}\rangle+\bar{\chi}_{ES}\langle c^\dagger c\rangle -\bar{\kappa}\langle \frac{p}{p_0}\rangle\;\;.\nonumber
 \end{eqnarray}
 while the complete master equation in these rescaled parameters takes the form,
  \begin{equation}
 \frac{d\rho}{d\tau}=-i[a^\dagger a,\rho]+i\bar{\chi{_{ES}}}[(a^\dagger+a)c^\dagger c/\sqrt{2},\rho]/\hbar+{\cal D}[c^\dagger E_L]\rho+{\cal D}[cE_R]\rho+\kappa(\bar{N}+1){\cal D}[a]\rho +\kappa\bar{N}{\cal D}[a^\dagger]\rho\;\;.\label{mymaster1}
\end{equation}
where
\begin{equation}
E_{R/L}\equiv \sqrt{\bar{\gamma}_{R/L}}(c/c^\dagger)\exp\left[
\mp\lambda\delta\pm\lambda(a+a^\dagger)/\sqrt{2}\right]\;\;.
\end{equation}
In what follows we take the phonon bath to be at zero temperature
$\bar{N}=0$, and that the typical phonon bath excitation will have energies much smaller than that of the oscillator, i.e. below 1GHz .

We will, in particular, be interested in the ``semiclassical'' and
fully quantum steady-state behaviour of the right tunnel current, or
the expected value of the operator: \beq
 \bar{i}_R\equiv \frac{\omega_0}{e}i_R \equiv \bar{\gamma}_R e^{\mp 2\lambda \delta}  e^{+2\lambda  x/x_0 }  c^\dagger c= \bar{\gamma}_R e^{\mp 2\lambda \delta}  e^{+2\lambda  \bar{x}}  c^\dagger c \;\;,\label{IR}
\eeq where $\bar{x}\equiv x/x_0,\;\;\bar{p}\equiv p/p_0$, and
$\bar{i}_R$, is the re-scaled average current flowing from the
moving island to the drain. We wish to determine whether $\langle
\bar{i}_R(t\rightarrow\infty)\rangle=\langle \bar{i}_R^\infty
\rangle$, is {\em significantly} different in the two cases when the
endohedral spin is up or down. However it is not enough that the two
currents (endohedral spin up/down), are different but we must also
then determine the steady-state (or DC or zero-frequency), quantum
noise associated with these two steady-state currents. This DC noise
will allow us to determine how {\em distinguishable} the two current
signals are. This then will determine the minimum length of signal
acquisition time one can average over to achieve a signal to noise
ratio which will allow the endohedral spin states to be
distinguished to a high level of confidence.  In the following
subsections (\ref{subsection_semi}) we discuss the initial
``semi-classical'' and quantum mechanical steady-state for the
coupled electronic-vibronic shuttle system, paying particular regard
to the change in right-tunnel-current with endohedral spin-state.
This is done in a parameter range of the system where we are certain
that our numerics are very precise and faithfully simulate the
system.
%Hsi-Sheng: Gerard: Dian In subsection (\ref{Dian_Milburn}), we derive a master equation analytically, in the limit of large tunnel rates, for the vibronic system alone. We compare the vibronic  quantum steady-state resulting from this vibration-only master equation with the reduced vibronic steady-state of the simulated coupled electronic-vibronic system to again give us a further check of the accuracy of our numerics.
We find a parameter regime where  $\langle
\bar{i}_R^{\uparrow}\rangle/\langle \bar{i}_R^\downarrow\rangle$ can
be substantial, indicating that the endohedral spin-state
measurement could be possible. In subsection (\ref{trajectory}), we
expand the size of the numerical simulation to cover a much larger
range of parameters (see parameter set (B) in Table I), which models
a much larger electrostatic drive force $\bar{\chi}_{ES}$. This is
done using the method of quantum trajectories \cite{QTraj_history, carmichael, tan}.
This type of simulation is very costly and only a small number of
such simulations were performed. We found that when the
electrostatic driving force is increased, the ratio $\langle
\bar{i}_R^{\uparrow}\rangle/\langle \bar{i}_R^\downarrow\rangle$
remains  significant. Further the highly accurate quantum trajectory
simulations corresponds very well with the lower-resolution
steady-state results indicating that our steady-state numerics could
be trusted to quite large values of the electrostatic force. In
subsection (\ref{noise}), we derive the analytical expressions for
the noise spectra and numerically derive the DC-noise component over
a limited (but precisely modeled), range of parameters. We
reproduced the expected values for the DC noise of a standard,
spatially fixed quantum dot, in the case when $\lambda=0$, i.e. no
coupling between the electronic and vibration degrees of freedom. We
further found a slight decrease in this noise with the amount of
electrostatic driving and degree of the exponential coupling.
Finally, in subsection (\ref{averaging}), we gather together all
this information to estimate the measurement time required to
distinguish between the current signals due to endohedral spin
up(down).

\section{Numerical Analysis}
\subsection{Physical Parameters}
We now determine the typical values of the dimensionless parameters
appearing in the system's dynamics. We will take, $m_{\rm{C_{60}}}=
1.2\times 10^{-24}$kg, $\omega_0=2\pi\times 1$GHz, the potential
difference between the contacts to be $V=10$meV, and the spatial
separation between the contacts $D=10$nm. From this  we have
$x_0\sim 1.2$ \AA, $\delta\sim   0.18$, giving the physical
displacement $\bar{\delta}\sim .2$ \AA.  The dimensionless
displacement of the harmonically bound shuttle due to the
electrostatic force alone in units of $x_0$ is $\delta_{ES}\sim
|F_{ES}|/(m\omega^2_0x_0)\sim 28$, while $\lambda\sim 0.4$. From
symmetry conditions one might suspect that we cannot gain any
advantage from the exponential dependence of the tunneling
amplitudes on the shuttle-contact separation if the shuttle's
equilibrium position is exactly midway between the contacts
(effectively when $\bar{\gamma}_L=\bar{\gamma}_R$). Thus we choose
$\bar{\gamma}_L\sim 0.1$, while $\bar{\gamma}_R\sim 0.9$. We choose
$\bar{\kappa}\sim 2$, thus setting the system close to critical
damping. We also set $\bar{N}=0$, or assume a zero temperature
vibronic bath, throughout. The example re-scaled parameter values
that we study below are summarized in Table I.

\begin{table}[h!]
\begin{center}{\large
\begin{tabular}{|c|c|c|c|c|c|c|}
  \hline\hline
  & {\boldmath $\lambda$} & {\boldmath $\delta$} & {\boldmath $\bar{\chi}_{ES}$} & {\boldmath $\bar{\kappa}$} & {\boldmath $\bar{\gamma}_L$} & {\boldmath $\bar{\gamma}_R$  }\\ \hline\hline
 A & 0.4 & 0.18 & 28 & 2 & 0.1 & 0.9 \\ \hline
 B &  0.4 & 0.4 & 2 & 2 & 0.1 & 0.9 \\ \hline
 C & 0.4 & 0.4 & 10 & 2 & 0.1 & 0.8 \\ \hline \hline
\end{tabular}
} \caption{Parameter values for the model (A) typical physical
parameters, (B) initial model parameters studied in
``semiclassical'' and quantum steady state,  numerical analyses, (C)
parameters used in ``semiclassical'', low-resolution quantum
steady-state, and high-resolution quantum dynamical (quantum
trajectory), numerical analyses. }
  \end{center}\label{table1}
\end{table}

\subsection{Semi-Classical and Quantum Steady-State}\label{subsection_semi}
From Eq. (\ref{averagec}), one can obtain the steady-state solution
for the expectation values of $\bar{n}=\langle c^\dagger c\rangle$,
$\langle \bar{x}\rangle$, and $\langle \bar{p} \rangle$, to be
\begin{eqnarray}
\expect{\bar{n}}_\infty&=&\frac{1}{\gamma_R/\gamma_L e^{4\lambda \bar{x}_\infty \pm 4 \lambda \delta}} , \label{e:fpn}\\
\expect{\bar{x}}_\infty &= & \frac{\bar{\chi}_{ES}}{[1+(\kappa/2)^2]}\bar{n}_\infty , \label{e:fpx}\\
\expect{\bar{p}}_\infty &  = & \kappa \bar{n}_\infty . \label{e:fpy}
\end{eqnarray}

Thus the fixed point must satisfy:
\begin{equation}
\bar{x}_\infty=\frac{\bar{\chi}_{ES}}{1+(\kappa/2)^2}
\biggl(\frac{1}{\gamma_R/\gamma_L e^{4\lambda \bar{x}_\infty \pm 4
\lambda \delta}}\biggr) .
\end{equation}

Following the derivation outlined in \cite{utami}, the stability
of the fixed point can be found by looking at the eigenvalues derived from
a linearised dynamics. From \cite{utami},  the system 
possesses one real eigenvalue and two complex conjugate eigenvalues which
indicate the existence of a limit cycle. This limit cycle appears at
a bifurcation of the fixed point which happens when the eigenvalues
are purely imaginary. This condition appear at a critical value of
$\chi_{ES}$:
\begin{equation}
\chi_{ES}\equiv\chi_h = \frac{(A_*^2 + A_* \kappa + (\frac{\kappa}{2})^2 +1)A_*
\kappa}{4 \lambda \gamma_L \gamma_R} \label{e:chicritical}
\end{equation}
where $A_* = \gamma_L e^{-2 \lambda \bar{x} \mp 2 \lambda \delta} +
\gamma_R e^{2 \lambda \bar{x} \pm 2 \lambda \delta}$.

%%%%DW: Shall wel put in here our result of semiclassical near the bifurcation
%%%point?
% JT: Lets leave this completely aside...we have done no real serious study of what might happen here and we might be completely off. Rather than point people up avenues where we might go and have not really explored yet lets leave it out.

Interestingly, the fixed point for the two spins are different and
will result in a slightly different critical $\chi_h$ values. However we choose to operate away from  this regime  in order to again avoid any possibility of noise upsetting the distinguishability of the resulting signals.
% JT: inserted by DW, Commented out by JT:
%Here
%we are presented with two possibilities.  We can use this difference
%to differentiate the two signal resulting from the two spins if we
%can choose a value of $\chi_{ES}$ that lies between the two critical
%values. The behaviour of the current will then be different for the
%spin up and the spin down measurement.  This case will be presented
%somewhere else \cite{futurepaper}.

A more straight forward method is to keep system such that it falls
within the fixed point regime and proceed in calculating the
difference in the steady state current. This is possible by a
careful choice of values such that the coupling $\chi_{ES}$ are
always below this critical values for both of the spins.  This puts
the system well in the supercritical regime.  We can thus proceed by
looking at the steady state at the fixed point and thus eliminating
the possible error related to the existence of a stable limit cycle
at the outer position of the doped fullerene's oscillations.

From (\ref{e:fpn})-(\ref{e:fpy}), we see that when $\lambda=0$, there is obviously
no dependence of the current on the endohedral spin-state when
$\bar{\gamma}_R=\bar{\gamma}_L$. This is not the case when the
coupling is non-vanishing i.e. $\lambda>0$. This reflects the
inherent asymmetry of the current transport from the source to the
drain which is now mediated by an island which has, essentially,
steady-state tunneling probabilities to either contact which are now
effected by the endohedral spin dependent equilibrium position of
the island. Taking, the steady-state semiclassical right tunnel
current to be $\langle
\bar{i}_R\rangle_\infty^{SC}=\bar{\gamma}_R\exp(\pm
2\delta\lambda)\langle \bar{n}\rangle_\infty$, we have \beq
\frac{\langle \bar{i}_R^\uparrow\rangle_\infty^{SC}}{\langle
\bar{i}^\downarrow_R\rangle_\infty^{SC}}=\frac{ \gbl
e^{+2\lambda\delta}+\gbr e^{-2\lambda\delta}}{\gbl
e^{-2\lambda\delta}+\gbr e^{+2\lambda\delta}}\sim
e^{-4\lambda\delta}\;\;, \eeq in the limit of large $\gbr/\gbl$ ($\bar{\Gamma}_R/\bar{\Gamma}_L \gg e^{+ 4\lambda\delta}$), and
large $\lambda\delta$. Thus we might expect an exponential
dependance of the resulting current ratio on the endohedral
spin-state.

\begin{figure}[h!]
\includegraphics[width=100mm]{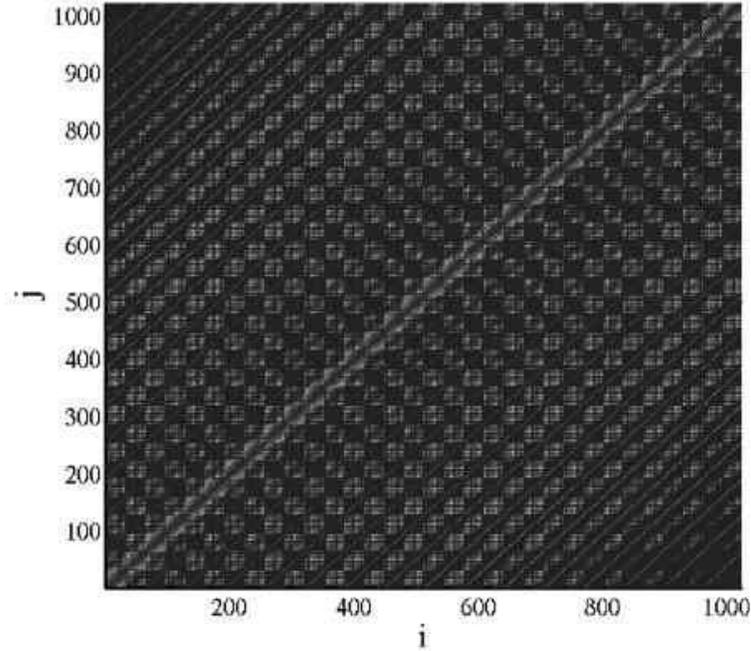}
\caption{Sparsity of the Liouvillian superoperator. Plot of
$\log_{10}$ of $|\rho_{ij}|$, when expanded into a  $4N^2\times
2N^2,\;\;N=16$ matrix. The steady-state solution is the eigenvector
with vanishing eigenvalue of this matrix. } \label{sparse_pic}
\end{figure}

We can obtain the steady-state solution to (\ref{mymaster1}), by
converting the resulting Master equation into a $4N^2\times 4N^2$,
(where $N$ is the size of the truncation of the Fock-state
representation of the vibronic Hilbert space), complex matrix and
searching for the eigenstate ($\rho_\infty$), with vanishing
eigenvalue. We use the inverse power method to find the eigenvector
with smallest eigenvalue \cite{golub}. In the following we mostly
choose $N=36$, giving matrices of size $4096\times 4096\sim
256$MBytes in memory. As the memory resources increase with $~N^4$,
we cannot realistically go beyond $N\sim 60$. However, if the
resulting large matrix is sparse then routines which only use matrix
vector products can be used to determine the steady-state solution
\cite{flindt}. In Figure \ref{sparse_pic}, we see that although
the matrix has many small elements, it is unclear that it is highly
sparse. As the current (\ref{IR}), is exponential in $\bar{x}$, the
expectation value $\langle \bar{i}_R\rangle$, could depend on very
small matrix elements. To ensure convergence we will work with the
full matrix in what follows and choose to truncate the bosonic
vibrational Hilbert space at $N=16$, or $N=32$.

\begin{figure}[h!]
 \begin{center}
 \setlength{\unitlength}{1cm}
\begin{picture}(6,12)
\put(-4,6.5){\includegraphics[width=76mm]{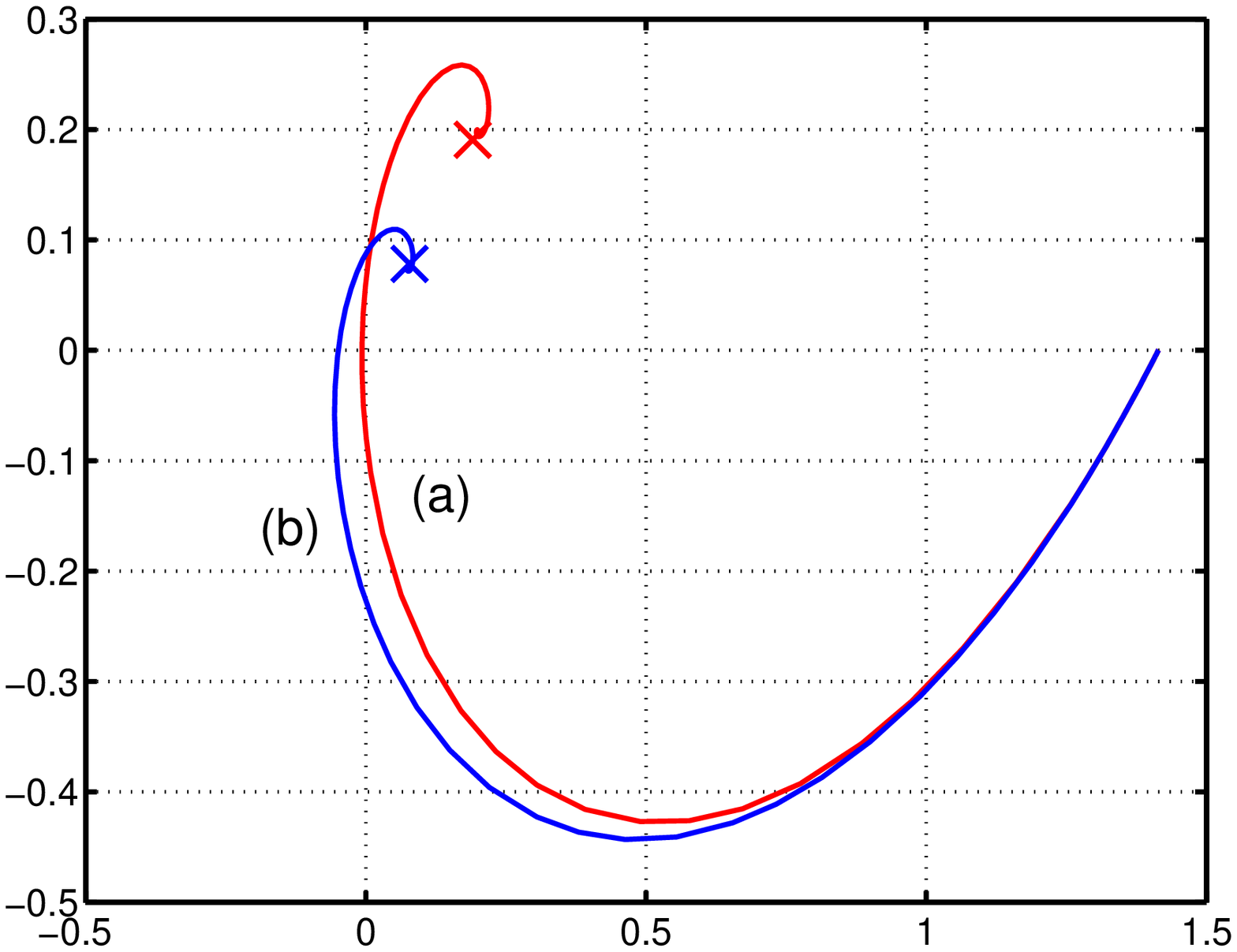}}
\put(4.5,6.5){\includegraphics[width=70mm]{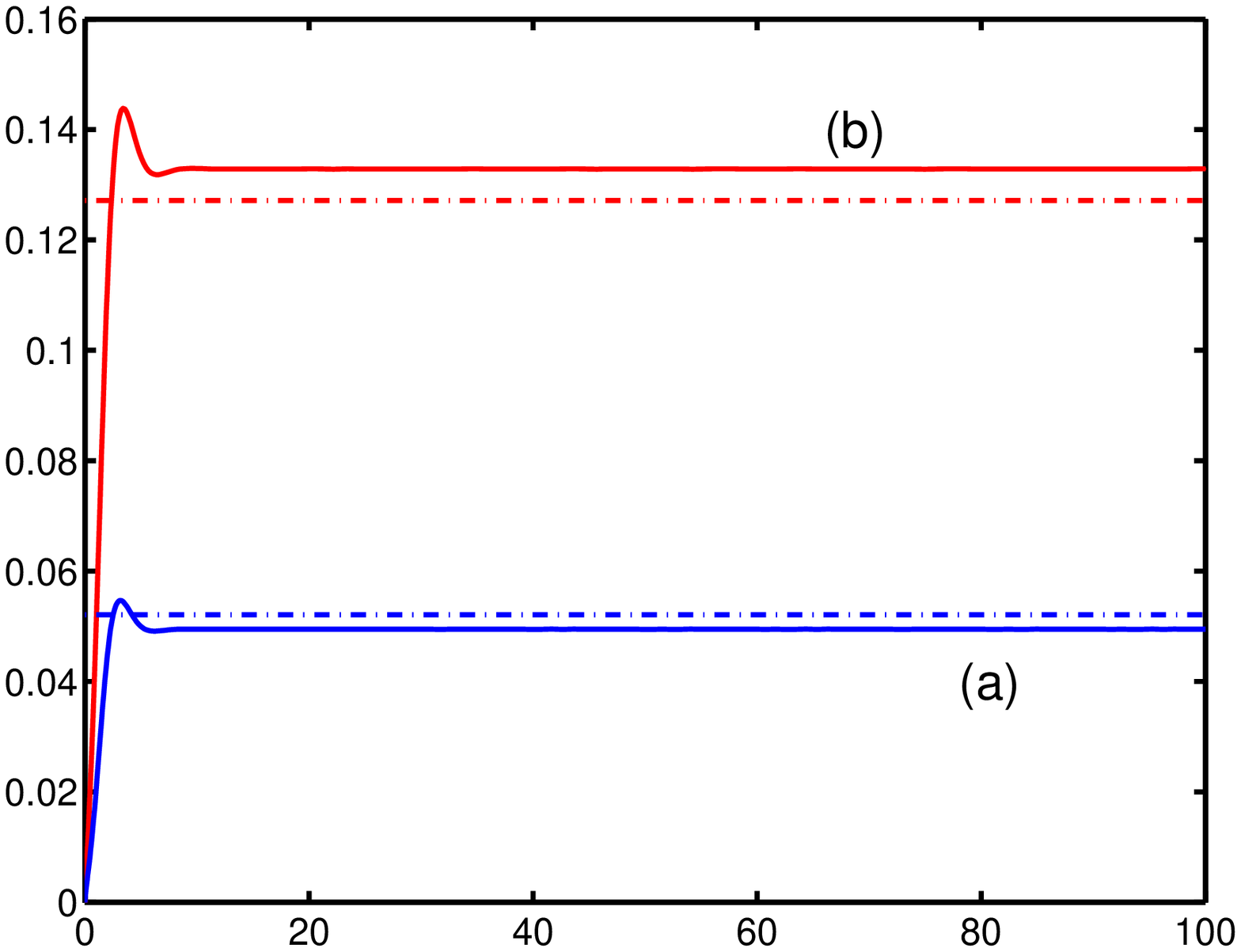}}
\put(0,.2){\includegraphics[width=70mm]{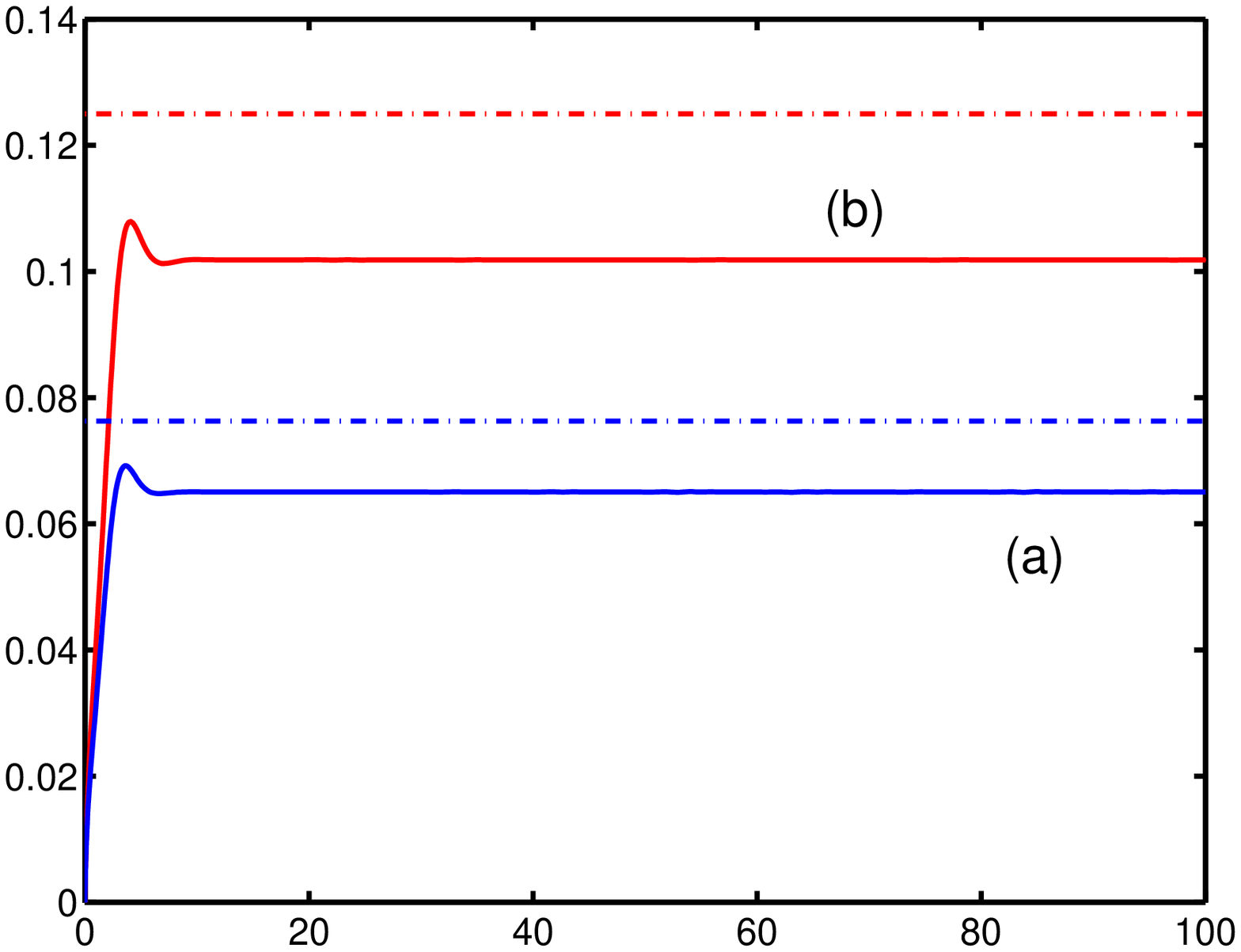}}
\put(0,6.1){\large$\langle \bar{x}\rangle$}
\put(8,6.1){\large$\tau$} \put(3.5,-.2){\large$\tau$}
\put(-3.8,9.1){\large\begin{rotate}{90}$\langle
\bar{p}\rangle$\end{rotate}}
\put(4.3,9.1){\large\begin{rotate}{90}$\langle c^\dagger
c\rangle$\end{rotate}} \put(-.2,3){\large\begin{rotate}{90}$\langle
\bar{i}_R\rangle$\end{rotate}} \put(-3.9,12){\Large (A)}
\put(4.1,12){\Large (B)} \put(-.7,5.5){\Large (C)}
\end{picture}
\end{center}
\caption{Semiclassical and steady-state quantum expectation values
for the case
$\bar{\chi}_{ES}=3,\;\;\bar{\kappa}=1,\;\;\gbl=.1,\;\;\gbr=.9,$ and
$\lambda=\delta=.4$. (A) semiclassical phase space evolution (in
displaced picture), (a)/(b) correspond to spin up/down or
$\pm\delta$; (B) graphs of $\langle c^\dagger c\rangle$, for the two
endohedral spin states (a) and (b), and where solid curves are the
semiclassical solutions and the dashed are the quantum steady-state
expectation values with $N=32$; (C) graphs of $\langle
\bar{i}_R\rangle$, for the two endohedral spin states (a) and (b),
and  where solid curves are the semiclassical solutions and the
dashed are the quantum steady-state expectation values with $N=32$.}
\label{semicheck1}
\end{figure}

\begin{figure}[h!]
\setlength{\unitlength}{1cm}

\begin{center}
\begin{picture}(6,9)
\put(-4,0){\includegraphics[scale=.8]{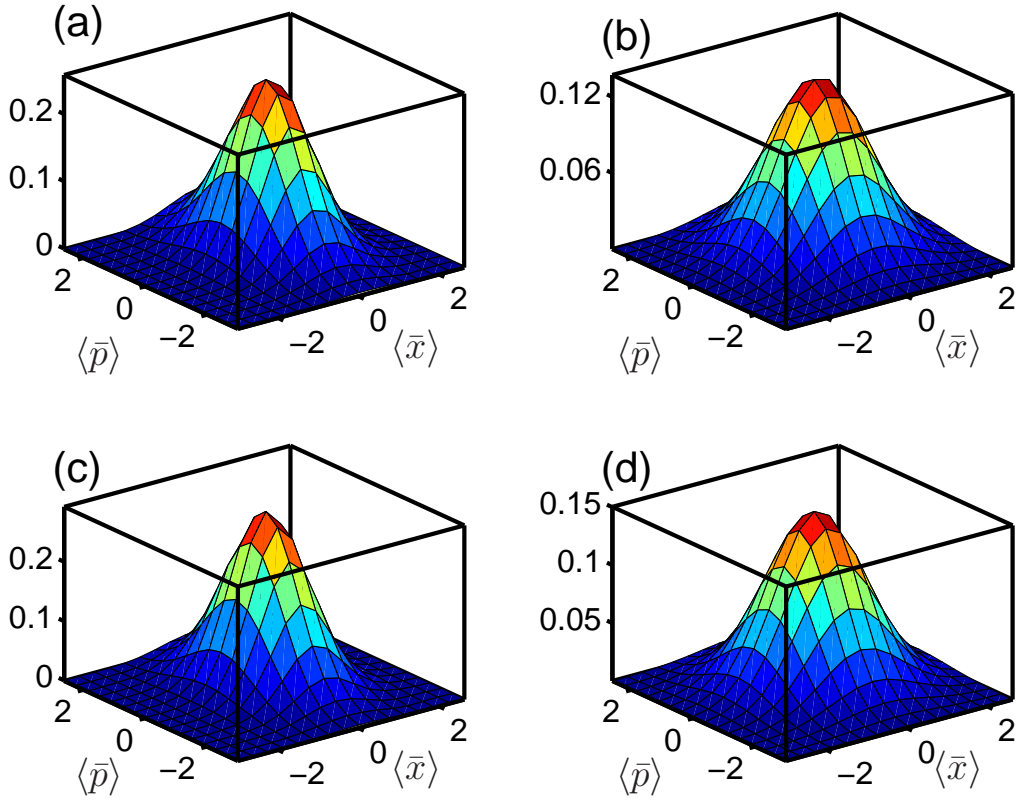}}
\put(-3.1,0.1){\Large$\langle \bar{p}\rangle$}
\put(1.1,0.2){\Large$\langle \bar{x}\rangle$}

\put(4.1,0.1){\Large$\langle \bar{p}\rangle$}
\put(8.3,0.2){\Large$\langle \bar{x}\rangle$}

\put(-3.1,5.7){\Large$\langle \bar{p}\rangle$}
\put(1.1,5.8){\Large$\langle \bar{x}\rangle$}

\put(4.1,5.7){\Large$\langle \bar{p}\rangle$}
\put(8.3,5.8){\Large$\langle \bar{x}\rangle$}

\end{picture}
\end{center}
\caption{Plot of the Quasi-Distribution functions for the quantum
steady-state, (a) and (c) are Wigner functions of the reduced
vibronic steady-state, $\rho_\infty^{vib}={\rm
Tr}_{dot}[\rho_\infty]$, for the cases of endohedral spin up  and
down; (b) and (d) are Husimi function for same. }
\label{quasiWQcheck}
\end{figure}

\begin{figure}[h!]
 \begin{center}
 \setlength{\unitlength}{1cm}
\begin{picture}(6,12)
\put(-4,0){\includegraphics[width=140mm]{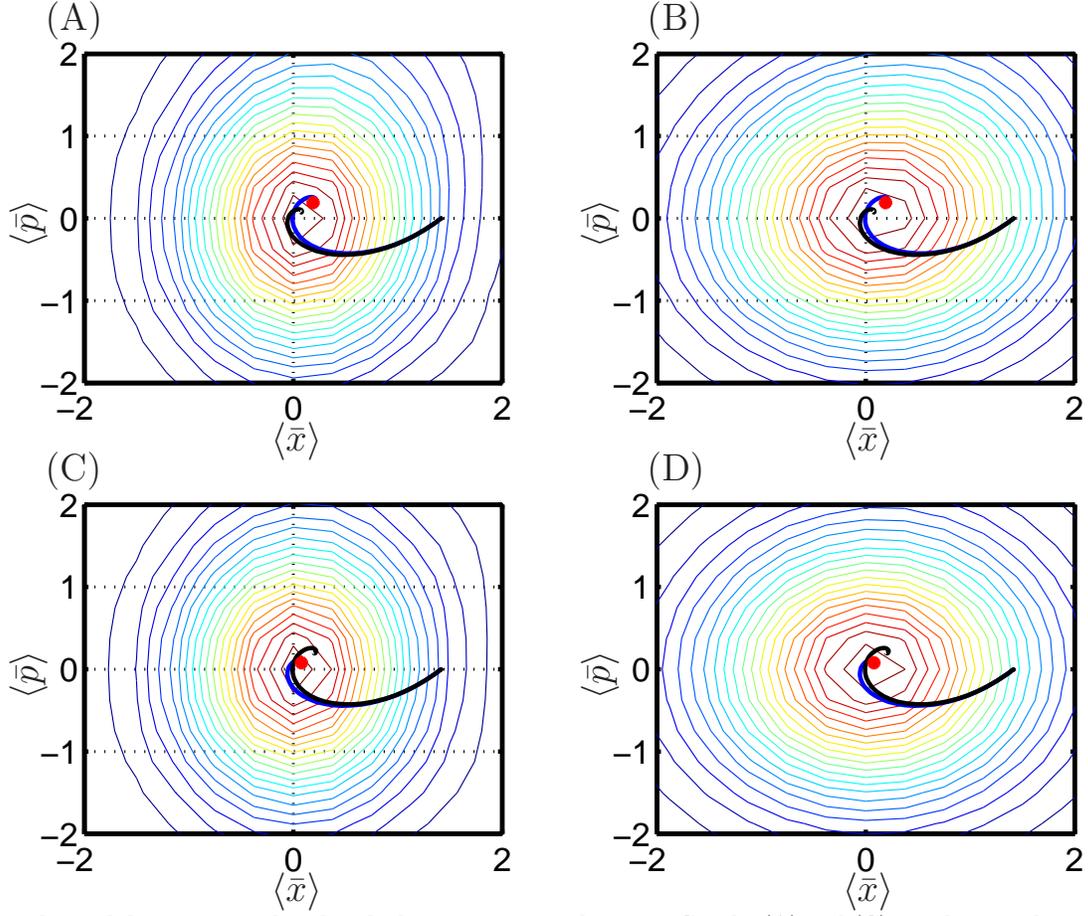}}
\put(-.9,5.7){\Large$\langle \bar{x}\rangle$}
\put(6.7,5.7){\Large$\langle \bar{x}\rangle$}
\put(-.9,-.3){\Large$\langle \bar{x}\rangle$}
\put(6.7,-.3){\Large$\langle \bar{x}\rangle$}

\put(-4,2.4){\large\begin{rotate}{90}\Large$ \langle \bar{p}\rangle
$\end{rotate}} \put(3.6,2.4){\large\begin{rotate}{90}\Large$ \langle
\bar{p}\rangle $\end{rotate}}

\put(-4,8.4){\large\begin{rotate}{90}\Large$ \langle \bar{p}\rangle
$\end{rotate}} \put(3.6,8.4){\large\begin{rotate}{90}\Large$ \langle
\bar{p}\rangle $\end{rotate}}

\put(-3.9,11.3){\Large (A)} \put(4.1,11.3){\Large (B)}
\put(-3.9,5.3){\Large (C)} \put(4.1,5.3){\Large (D)}
\end{picture}

\caption{Semiclassical dynamics overlayed with the quantum
steady-state. Graphs (A) and (C) are the semiclassical (blue/black
curves), quantum steady-state phase point $( \langle
\bar{x}\rangle_\infty, \langle \bar{p}\rangle_\infty)$, as a red
symbol, and the contours of the quantum steady-state  Wigner
function, for the case of endohedral spin up and down. The blue
curve is the appropriate semiclassical phase-space trajectory for
the spin up(down), while the black curve is the other case shown for
comparison. Graphs (B) and (D) are same except the Husimi function
is used.}
\end{center}\label{quasisemicheck}
\end{figure}

In Figure\ref{semicheck1}, we plot the ``semiclassical''
evolution of (\ref{averagec}), and quantum steady-state expectation
values when
$\bar{\chi}_{ES}=3,\;\;\bar{\kappa}=1,\;\;\gbl=.1,\;\;\gbr=.9,$ and
$\lambda=\delta=.4$. We see that the different endohedral spin
states lead to very different phase space dynamics, differing island
populations $\langle \bar{n}\rangle$, and currents $\langle
\bar{i}_R\rangle_\infty$. One further notices a remarked difference
between the magnitudes of the quantum and semiclassical expectation
values for the currents in Fig. (\ref{semicheck1}.c). This
difference grows with the size of $\bar{\chi}_{ES}$, and is
primarily due to the extended nature of the quantum state in the
vibronic phase space. With the availability of the full quantum
steady-state solution we are also able to visualise the quantum
state via quasi-distribution functions and in Figure
\ref{quasiWQcheck}, we see that the quantum state is localised
near the phase space origin. Moreover, from the compactness of the
resulting Wigner or Husimi representations we can be relatively
certain that the quantum steady-state solution is well represented
with a Fock state truncation level set at $N=32$. As
$\bar{\chi}_{ES}$ is increased towards the more physically realistic
regime of $\bar{\chi}_{ES}\sim 20$,  one is justifiably concerned
whether a truncation level of $N\sim 32$, would remain sufficient to
accurately represent the resulting quantum steady-state solution. We
will see below however, that $N\sim 32$, remains fairly accurate
when $\bar{\chi}_{ES}=10$. In the section below we are primarily
interested in knowing with relative certainty that our quantum
simulations are highly accurate. In Figure \ref{quasisemicheck},
we superimpose the semiclassical phase space evolution on to the
Wiger function of the quantum steady-state solution for the two
cases of endohedral spin up and down. We note that due to a slight
assymetry in the Wigner function the quantum expectation value for
the steady-state equilibrium solution is not located at the peak of
the Wigner function.

\begin{figure}[h!]
 \begin{center}
 \setlength{\unitlength}{1cm}
\begin{picture}(6,8)
\put(-2,0){\includegraphics[width=100mm]{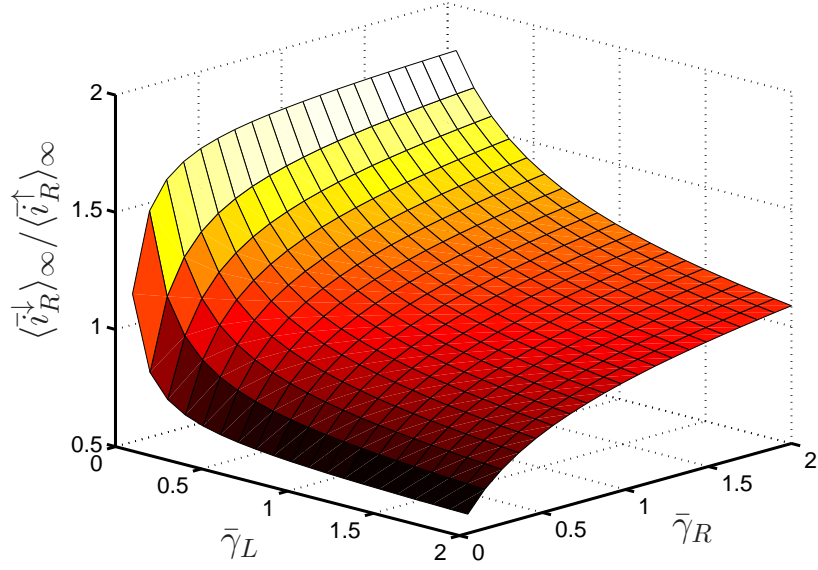}}
\put(0,.2){\Large$\gbl$} \put(6,.4){\Large$\gbr$}
\put(-2.3,3){\begin{rotate}{90}\Large$\langle
\bar{i}_R^{\downarrow}\rangle_\infty/\langle
\bar{i}_R^\uparrow\rangle_\infty$\end{rotate}}

\end{picture}

\caption{Graph of $\langle
\bar{i}_R^{\downarrow}\rangle_\infty/\langle
\bar{i}_R^\uparrow\rangle_\infty$, computed via the quantum
steady-state for a range of values of the left and right tunnel
amplitudes, $\gbl$ and $\gbr$. For large $\gbr/\gbl$, the ratio
tends to $\exp(+4\lambda\delta)$.}
\end{center}\label{quantcurrentratioshi}
\end{figure}

We can see from this analysis that there is a significant difference
between $\langle \bar{i}_R^{\uparrow}\rangle$, and $\langle
\bar{i}_R^\downarrow\rangle$. To characterize this difference more
systematically we numerically evaluate the ratio $\langle
\bar{i}_R^{\uparrow}\rangle/\langle \bar{i}_R^\downarrow\rangle$, as
a function of $\gbl$ and $\gbr$. We choose
$\bar{\chi}_{ES}=3,\;\;\bar{\kappa}=2,\;\;\lambda=\delta=.4$, and
range through values of $\gbl$ and $\gbr$, computing the resulting
quantum steady-state current ratios for the two cases of endohedral
spin up and spin down. The results are shown in Figure
\ref{quantcurrentratioshi}, and in the limit of large $\gbr/\gbl$ ($\bar{\Gamma}_R/\bar{\Gamma}_L \gg e^{+ 4\lambda\delta}$),
the current ratio appears to asymptote, as it should to
$\exp(4\lambda\delta)\sim 1.9$. By repeating these simulations at
the reduced truncation $N=16$, and finding that the resulting
current ratios are within $10^{-9}$ of those computed at $N=32$,
indicates that our quantum steady-state simulations are extremely
precise.

%\subsection{Milburn/Goan/Dian Section}

\subsection{Quantum Trajectories}\label{trajectory}
As mentioned above, going much beyond a Fock state truncation of
$N\sim 60$, with the full matrix representation of the Liouvillian
is not possible without exotic computational resources. Another
method of simulating the dynamical evolution of the full quantum
master equation (\ref{mymaster1}), is to use the method of {\em
quantum trajectories} \cite{QTraj_history, carmichael, tan}. Briefly, in this method
one represents the quantum master equation (\ref{mymaster1}), as an
average of a stochastic Schr\"{o}edinger equation over some
stochastic noise. The temporal evolution of an initial state
determined by this stochastic Schr\"{o}edinger equation is known as
a quantum trajectory. One can determine the time evolution of
quantum expectation values of an observable by taking the
expectation values of that observable within a single quantum
trajectory and then averaging the expectation values found over
numerous stochastic repetitions of the quantum trajectory. Since the
method primarily involves the Schr\"{o}edinger evolution of pure
quantum states, the quantum evolution of an individual quantum
trajectory can be simulated with much higher Fock state truncations
$N$, than in the above steady-state method. The down side is that to
obtain precise expectation values one must average over a large
number of repetitions of the quantum trajectory.

\begin{figure}[h!]
\begin{center}
\setlength{\unitlength}{1cm} \linethickness{2pt}
\begin{picture}(6,9)
\put(-4,6){\includegraphics[width=15cm,height=3cm]{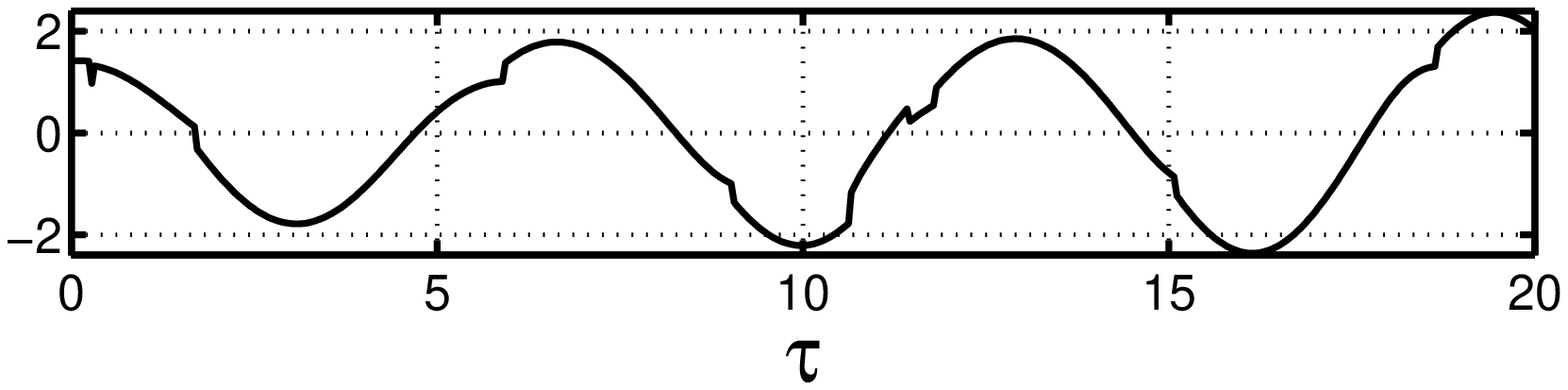}}
\put(-4.1,3){\includegraphics[width=15.1cm,height=3cm]{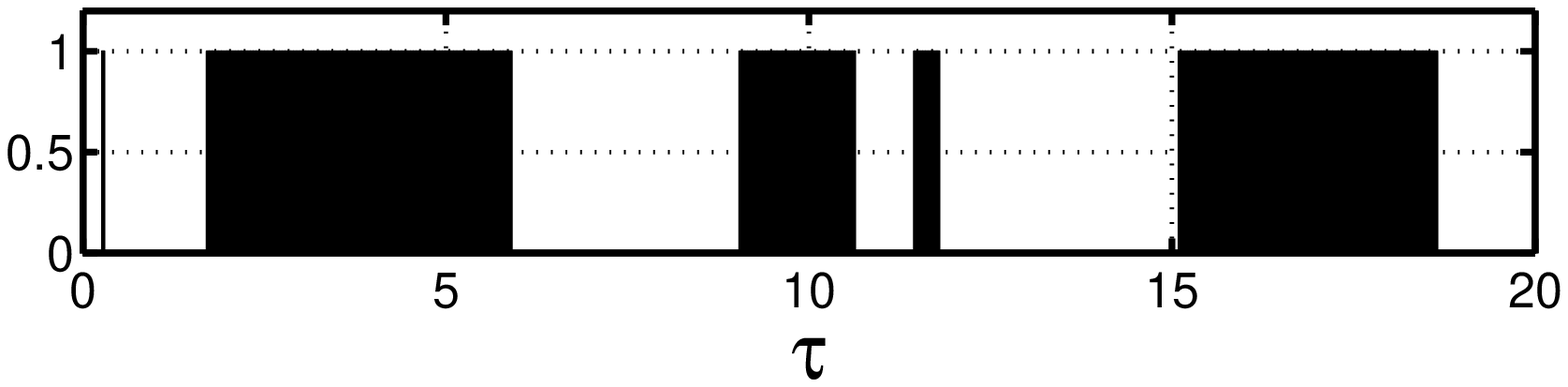}}
\put(-3.7,0){\includegraphics[width=14.7cm,height=3cm]{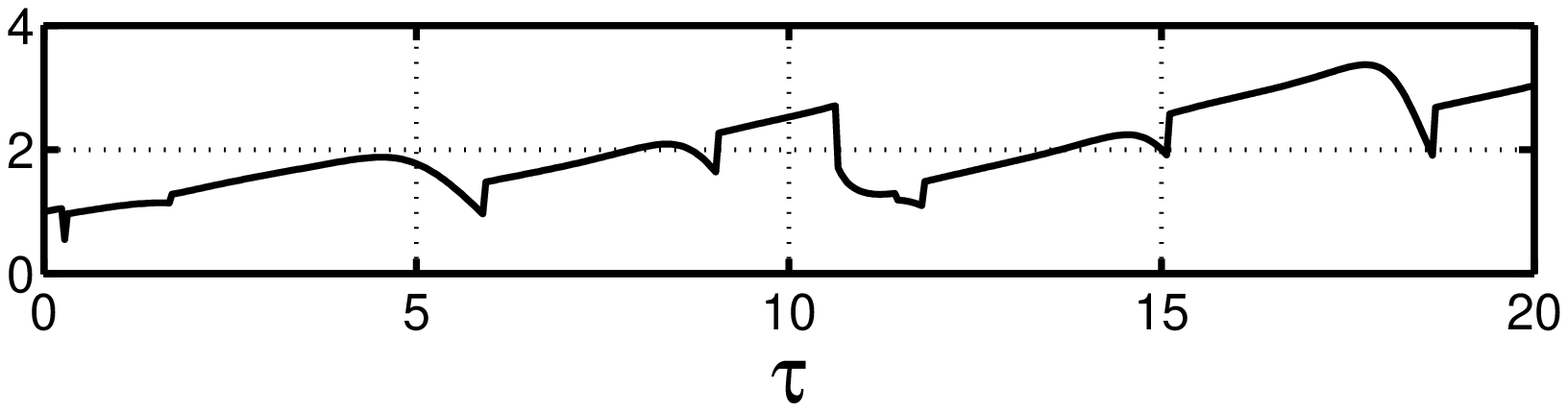}}
\put(-6,7.5){{\Large (a)}} \put(-6,4.5){\Large (b)}
\put(-6,1.5){\Large (c)} \put(-4.3,7.4){
\begin{rotate}{90}\large$\langle x/x_0\rangle$\end{rotate}}
\put(-4.3,4.4){ \begin{rotate}{90}\large$\langle c^\dagger
c\rangle$\end{rotate}} \put(-4.3,1.4){
\begin{rotate}{90}\large$\langle a^\dagger a\rangle$\end{rotate}}
\put(-2.18,9.5){\line(0,-1){.5}}\put(-2.18,7){\line(0,-1){1}}\put(-2.18,4){\line(0,-1){1.2}}\put(-2.18,1){\line(0,-1){.5}}
\put(0.8,9.5){\line(0,-1){.5}}\put(0.8,  7){\line(0,-1){1}}\put(.8,
4){\line(0,-1){1.2}}\put(.8,    1){\line(0,-1){.5}}
\put(3,9.5){\line(0,-1){.5}}\put(3,  7){\line(0,-1){1}}\put(3,
4){\line(0,-1){1.2}}\put(3,    1){\line(0,-1){.5}}
\put(4.05,9.5){\line(0,-1){.5}}\put(4.05,
7){\line(0,-1){1}}\put(4.05,    4){\line(0,-1){1.2}}\put(4.05,
1){\line(0,-1){.5}} \put(-2.31,0){\Large A} \put(.6,0){\Large B}
\put(2.8,0){\Large A} \put(3.88,0){\Large B}
\end{picture}
\end{center}
\caption{Quantum trajectories simulation (1 trajectory with a basis
of 100 Fock states), of  (\ref{mymaster}), for the case when
$\bar{\gamma}_R=\bar{\gamma}_L=1$, $\lambda=\delta=0.4$,
$\bar{\kappa}=\bar{n}=\delta=\eta=\bar{\chi}_{ES}=0$. Graph (a)
$\langle \bar{x}/x_0\rangle=\langle \bar{x}\rangle$ vs. $\tau$, (b)
$\langle c^\dagger c\rangle$, (c) $\langle a^\dagger a\rangle$. See
text for description of quantum jumps at times $A$ and $B$. }
\label{single_traj}
\end{figure}

The evolution of the stochastic Schrodinger equation corresponding
to (\ref{mymaster1}), when $\bar{N}=0$, is conditioned on the
appearance of random quantum jumps. The information gained by the
environment, in a sense, yields information which conditions the
quantum state following the quantum jump. In (\ref{mymaster1}),
there are three types of quantum jump: (I) $|\psi\rangle\rightarrow
c^\dagger E_L|\psi\rangle$ (unnormalised), which corresponds to an
electron tunneling onto the island from the Fermi bath of the left
electrode with some associated monitoring (via $E_L$), of the
island's position state by the vibrational bath, (II)
$|\psi\rangle\rightarrow c E_R|\psi\rangle$, corresponding to an
electron moving off the island into the Fermi bath of the right
electrode with some associated monitoring (via $E_R$), of the
island's position state by the vibrational bath, and (III)
$|\psi\rangle\rightarrow a|\psi\rangle$, where a motional quanta
escape's from the island's position state into the vibrational bath.
In Figure \ref{single_traj}, we show an example of a single
quantum trajectory of the coupled electronic-vibronic shuttle system
corresponding to (\ref{mymaster1}). We choose no motional damping
and set most of the parameters to vanish so we can illustrate the
effects of the quantum jumps clearly.  At time points (A) in Figure
\ref{single_traj}, a single electron jumps onto the island. Since
this is more likely for negative $\bar{x}$, due to the exponential
dependance of $E_L$ on position, we see a jump in the island's
position towards the source contact located in the $-\bar{x}$
region. At time points (B), this electron now jumps off the island
into the drain. Since again this is more likely closer to the drain
due to the form of $E_R$, we see a quantum jump in the island's
position to larger values of $+\bar{x}$. While the island is either
empty/occupied, the non-Hermitian evolution suffered by the quantum
trajectory causes the system to move towards the situation where the
occupied/empty state becomes more and more probable. As a
consequence of the exponential dependances of $E_L$ and $E_R$, this
means that the oscillations grow towards larger $+\bar{x}$, if the
island is occupied, while they grow towards $-\bar{x}$, if the
island is empty. The net effect is to pump the vibrational motion of
the shuttle during the single trajectory and even though this growth
is interspersed with dips due to the movement on or off of an
electron. Once averaged over many quantum trajectories, the
resulting effect is to pump the average energy of the oscillation in
an unbounded manner (if there is no motional damping).

\begin{figure}[h!]
\setlength{\unitlength}{1cm} \linethickness{1.7pt}
\begin{center}
\begin{picture}(6,12)
\put(-4,0){\includegraphics[scale=.8]{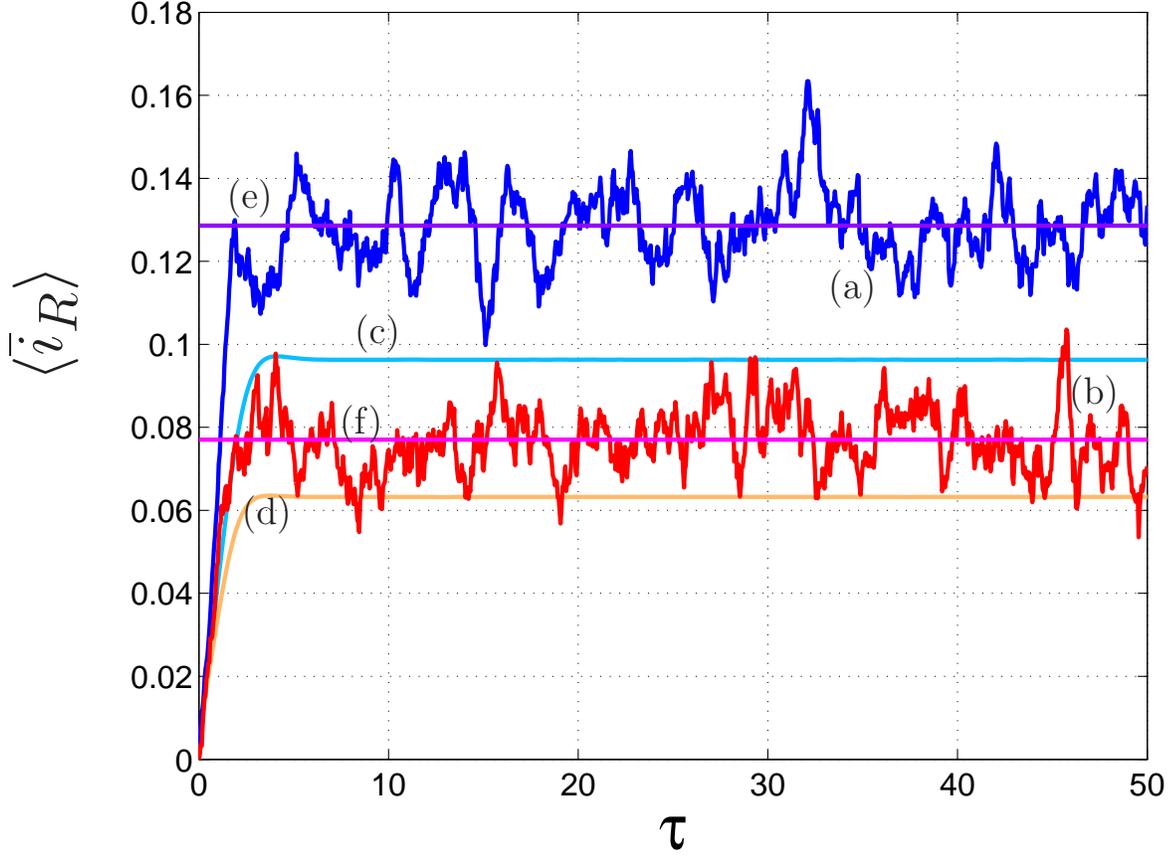}}
\put(-5,6.4){ \begin{rotate}{90}\Huge$\langle
\bar{i}_R\rangle$\end{rotate}} \put(5.3,7.5){{\Large
(a)}}\put(8.5,6.1){\Large (b)} \put(-1,6.9){\Large
(c)}\put(-2.5,4.5){\Large (d)} \put(-2.7,8.7){\Large
(e)}\put(-1.2,5.7){\Large (f)}
\end{picture}
\end{center}
\caption{Graphs of the right tunnel current given as a function of
rescaled time $\tau$, for the two spin-states ($\pm \delta$). Graphs
(a-b)/(blue-red) is $\langle \bar{ i}_R\rangle$ computed via a
quantum trajectories solution of (\ref{mymaster}),  for spin up/down
(2000 trajectories). Graphs (c-d)(turquoise-orange) are the
re-plotted ``semiclassical'' solutions for the two cases of the
shuttle spin state. Graphs (e-f)(magenta-purple) are $\langle \bar{
i}_R\rangle$ computed via the steady-state solution for
(\ref{mymaster}), on a $N=40$, Fock basis representation. Again we
have set
$\bar{\kappa}=2,\;\;\bar{\chi}_{ES}=10,\;\;\gamma_L=.1,\;\;\gamma_R=.8,\;\;\lambda=\delta=.4$.
} \label{traj_current}
\end{figure}

We now consider the full-blown quantum trajectory simulation of the system using the parameter values $\bar{\chi}_{ES}=10,\;\;\bar{\kappa}=2,\;\;\gbl=.1,\;\;\gbr=.8,$ and $\lambda=\delta=.4$. We now have chosen a far larger value for the electrostatic force and are able to expand the Fock basis truncation to $N=100$ using quantum trajectories. To determine the steady state quantum solution using  a full matrix representation would now require over 26GBytes of memory! The resulting simulation averages over 2000 quantum trajectories and required many days to execute on a 2GHz Operton Dual processor 64-bit computer. In Figure \ref{traj_current}, we plot the right tunnel current $\langle \bar{i}_R\rangle$, as computed via quantum trajectories for the two endohedral spin states. We also plot the ``semiclassical'' dynamical evolution and the steady-state quantum expectation values for the currents where the latter is computed at the necessarily lower Fock truncation of $N=40$. We can clearly see that the (incredibly computationally expensive), quantum trajectory simulation remarkably agrees very well with the values obtain from the steady-state numerics. It is not possible to plot out the Wigner function from the quantum trajectory simulation as this method can only yield the values of a small number of expectation values without running into the difficulty of requiring huge amounts of storage. However, the results shown in Figure \ref{traj_current}, strongly indicates that the technique of directly determining the quantum steady-state via the eigenvector of the Liouvillian matrix with vanishing eigenvalue,  can extend to very large values of the electrostatic driving force $\bar{\chi}_{ES}$, without any significant lack of precision. As the physically relevant regime for $\bar{\chi}_{ES}\sim 20$, the behaviours we obtain here with $\bar{\chi}_{ES}\sim 3-10$, should therefore also hold in the physically relevant  regime.  \\[20pt]

\subsection{Estimating the Steady-State Quantum Current Noise}\label{noise}
There are a number of methods to deduce the noise power spectral
density $S(\omega)$. We will use the methods outlined in
\cite{Goan235307}.

We first represent the instantaneous right tunnel current as
$i_R(t)=edN/dt$, where $dN(t)$, is a classical point process which
represents the number (either zero or one), of tunneling events seen
in an infinitesmal time, $dt$, and $e$ is the electric charge. We
can consider $dN(t)$ to be the increment in the number of electrons
$N(t)$, in the drain electrode in a time $dt$. The steady-state
current is computed by $i_R^\infty={\rm E}[i(t)]_{\infty}$, where
${\rm E}[\cdot]$, represents an average over the stochastic process
describing the tunneling events. In our case this is given by, \beq
i_R^\infty=\gamma_R\langle {\cal J}[c\,e^{+\bar{\lambda} x
}]\rho_\infty \rangle \;\;, \eeq where the jump superoperator ${\cal
J}[A]B\equiv ABA^\dagger$. The fluctuations in the current are
quantified by the two-time correlation function \beq
G(\tilde{t})\equiv {\rm
E}[i_R(t)i_R(t+\tilde{t})-i_R^{\infty\,2}]_\infty=ei_R^\infty\delta(\tilde{t})+\langle
i_R(t)i_R(t+\tilde{t})\rangle_\infty^{\tilde{t}\ne
0}\;\;.\label{mycorrel} \eeq From the theory of open quantum systems
\cite{carmichael}, one can show that \beq
G(\tilde{t})=ei_R^\infty\delta(\tilde{t})+e^2\left\{{\rm Tr}[{\cal
J}_R e^{{\cal L}\tilde{t}}{\cal J}_R\rho_\infty]-{\rm Tr}[{\cal
J}_R\rho_\infty]^2\right\}^\infty_{\tilde{t}\ne 0}\;\;, \eeq where
${\cal J}_R=\sqrt{\gamma_R}\,c\,\exp(-\bar{\lambda} x)$ (see
(\ref{myexponential})), and the Liouvilian evolution, $\exp( {\cal
L}\tilde{t})$, is according to the master equation (\ref{mymaster}).
In the case of a decoupled system ($\lambda=0$), this leads to a
normally ordered correlation function while one obtains an
antinormally ordered correlation function if one were studying the
noise of the left tunnel current \cite{milburn_ajp}. For the coupled
system the interpretation of the operator ordering is more
complicated.

The current noise spectral density is given by \beq S(\omega)\equiv
2 \int_{-\infty}^{+\infty}\,d\tilde{t} G(\tilde{t})\,
e^{i\omega\tilde{t}}\;\;. \eeq In terms of our rescaled variables
and parameters, $\tau=\omega_0 t$, and (\ref{rescaled}), we set
$\tilde{\tau}=\omega_0\tilde{t}$, to obtain, \beq
G(\tilde{\tau})=ei_R^\infty\delta(\tilde{\tau})\omega_0+e^2\omega_0^2\bar{\gamma}_R^2\left\{
{\rm Tr}[{\cal J}[ce^{+\lambda \bar{x}}] e^{{\cal L}\tilde{t}} {\cal
J}[+ce^{+\lambda \bar{x}}] \rho_\infty] -{\rm Tr}[ {\cal J}[c
e^{+\lambda \bar{x}}] \rho_\infty]^2\right\}^\infty_{\tilde{\tau}\ne
0}\;\;, \eeq where now the Liouvillian evolution is given by
(\ref{mymaster1}). The spectral density is now given by \beq
S(\omega/\omega_0)=2ei_R^\infty+e^2\omega_0\bar{\gamma}_R^2\int_{-\infty}^{+\infty}\,e^{i
(\omega/\omega_0)\tilde{\tau}}\,d\tilde{\tau}\, \left\{ {\rm
Tr}[{\cal J}[ce^{+\lambda \bar{x}}] e^{{\cal L}\tilde{t}} {\cal
J}[ce^{+\lambda \bar{x}}] \rho_\infty]-{\rm Tr}[ {\cal J}[c
e^{+\lambda \bar{x}}]\rho_\infty]^2\right\}^\infty_{\tilde{\tau}\ne
0}\;\;. \eeq

From this we can derive the Fano factor,
$F=S(\omega)/(2ei^\infty_R)$, to be \beq
F(\omega/\omega_0)=1+\frac{\bar{\gamma}_R^2}{2\bar{i}^\infty_R}
\int_{-\infty}^{+\infty}\,e^{i
(\omega/\omega_0)\tilde{\tau}}\,d\tilde{\tau}\, \left\{ {\rm
Tr}[{\cal J}[ce^{+\lambda \bar{x}}] e^{{\cal L}\tilde{t}} {\cal
J}[ce^{+\lambda \bar{x}}] \rho_\infty]-{\rm Tr}[ {\cal J}[c
e^{+\lambda \bar{x}}]\rho_\infty]^2\right\}^\infty_{\tilde{\tau}\ne
0}\;\;. \eeq

\begin{figure}[h!]
 \begin{center}
 \setlength{\unitlength}{1cm}
\begin{picture}(6,8)
\put(-2,0){\includegraphics[width=100mm]{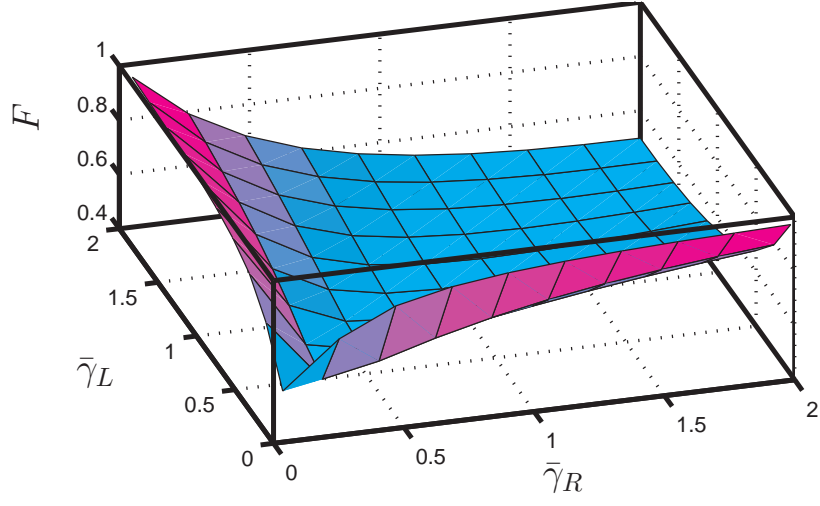}}
\put(-2,1.4){\Large$\gbl$} \put(4.2,0){\Large$\gbr$}
\put(-2.5,4.7){\begin{rotate}{90}\Large$F$\end{rotate}}

\end{picture}

\caption{Graph of Fano function $F$, for the decoupled system
$\lambda=0,\;\; \bar{\kappa}=.1,\;\;\bar{\chi}_{ES}=0$. Numerical
results agree with $F=(\gbl^2+\gbr^2)/(\gbl+\gbr)^2$.}
\end{center}\label{Fanocheck}
\end{figure}

The Fano factor $F$ gives information on the statistics of the
tunnel current noise. If $F=1$ then the noise is completely
uncorrelated/Poissonian, i.e. white noise. If $F>1$, then the noise
is known as super-Poissonian and the tunnel events are bunched. If
$F<1$, then the noise is known as sub-Poissonian and the tunneling
events are anti-bunched, i.e. there is little chance that a second
tunnel event will closely follow a previous event. Quantum
correlations within the coupled system are primarily responsible for
$F\ne 0$. In the case of the decoupled system, In Figure
\ref{Fanocheck}, we compare the results for the decoupled system
($\lambda=0$), with the standard results for a two-state sequential
tunneling process \cite{buttiker}, \beq
\bar{i}_R^\infty=\frac{\gbl\gbr}{\gbl+\gbr}\;\;,\qquad
F=\frac{\gbl^2+\gbr^2}{(\gbl+\gbr)^2}\;\;.\label{butt} \eeq

\begin{figure}[h!]
\begin{center}
\setlength{\unitlength}{1cm} \linethickness{2pt}
\begin{picture}(6,13)
\put(-4,11.2){\includegraphics[width=15cm,height=2.5cm]{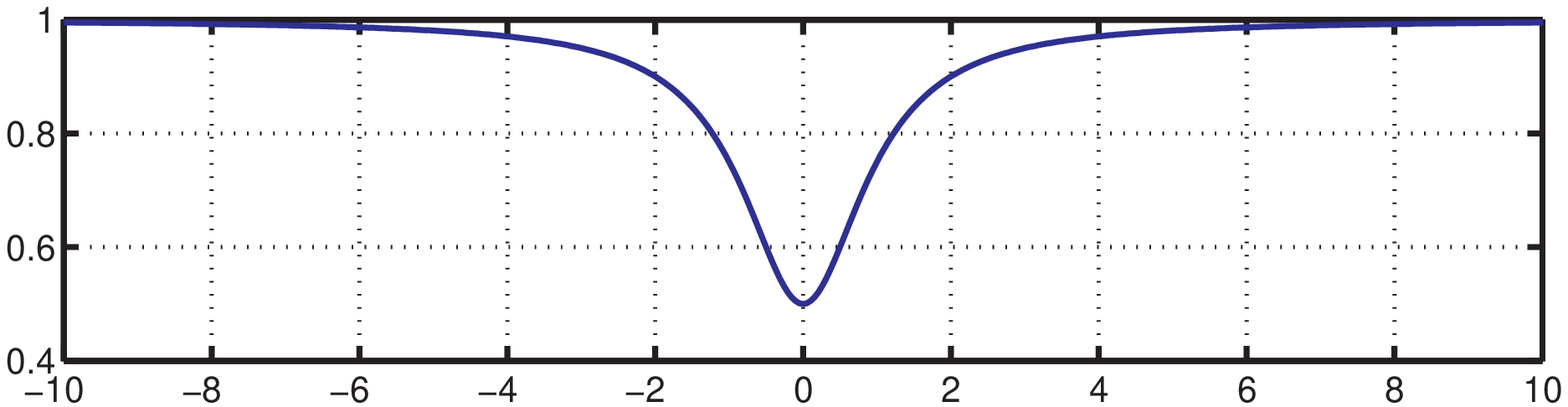}}
\put(-4.,8.4){\includegraphics[width=14.9cm,height=2.5cm]{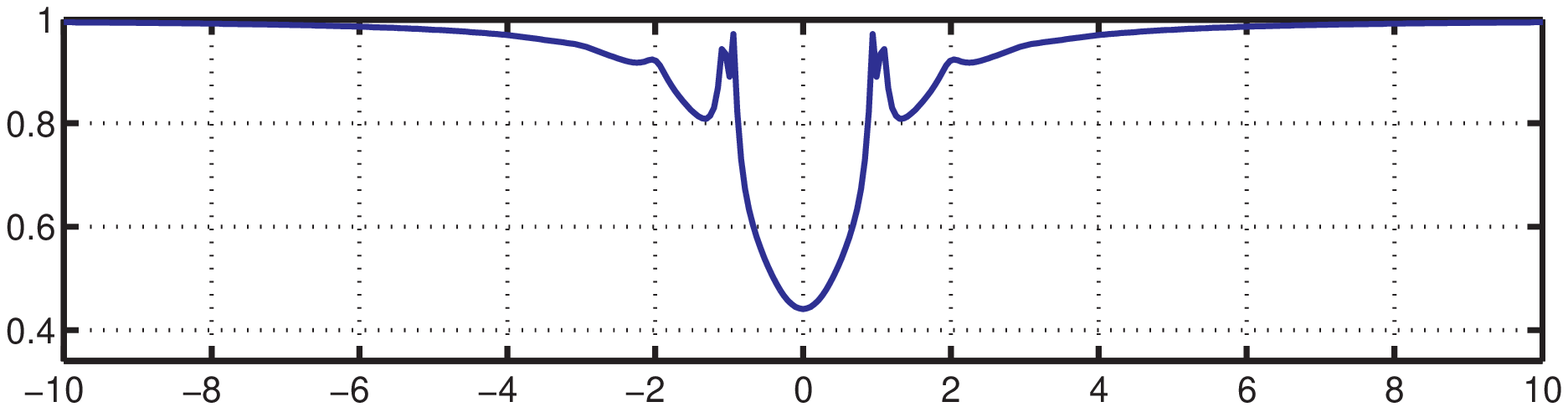}}
\put(-4.1,5.6){\includegraphics[width=15cm,height=2.5cm]{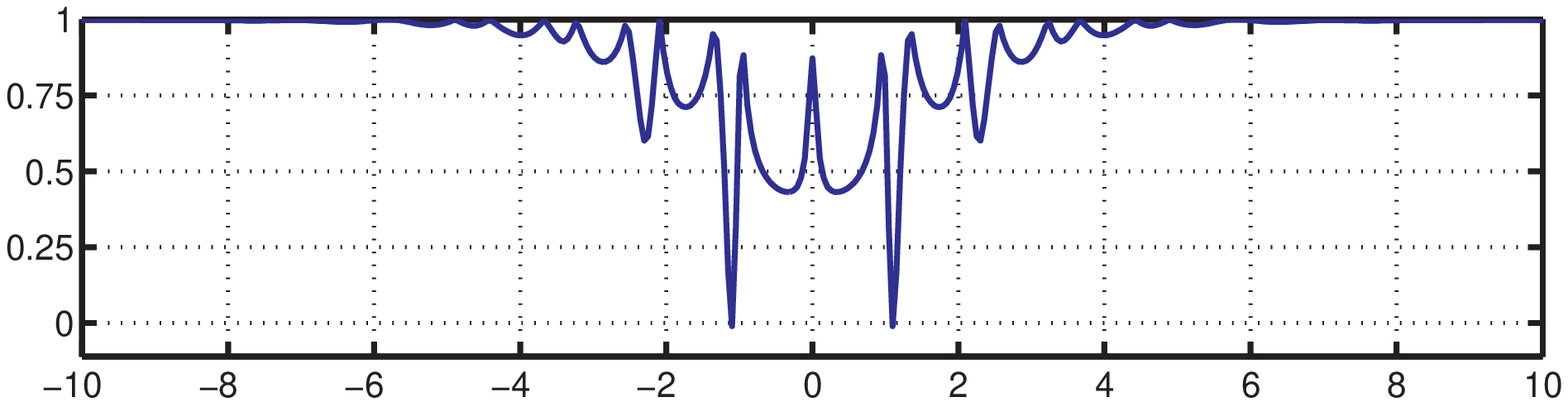}}
\put(-3.85,2.8){\includegraphics[width=14.8cm,height=2.5cm]{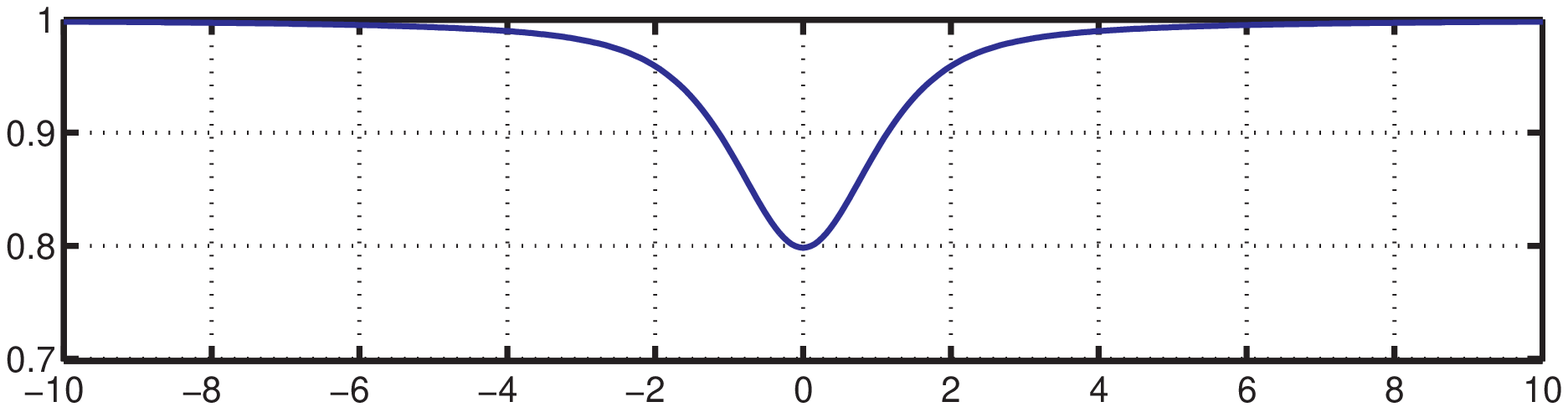}}
\put(-3.85,0){\includegraphics[width=14.8cm,height=2.5cm]{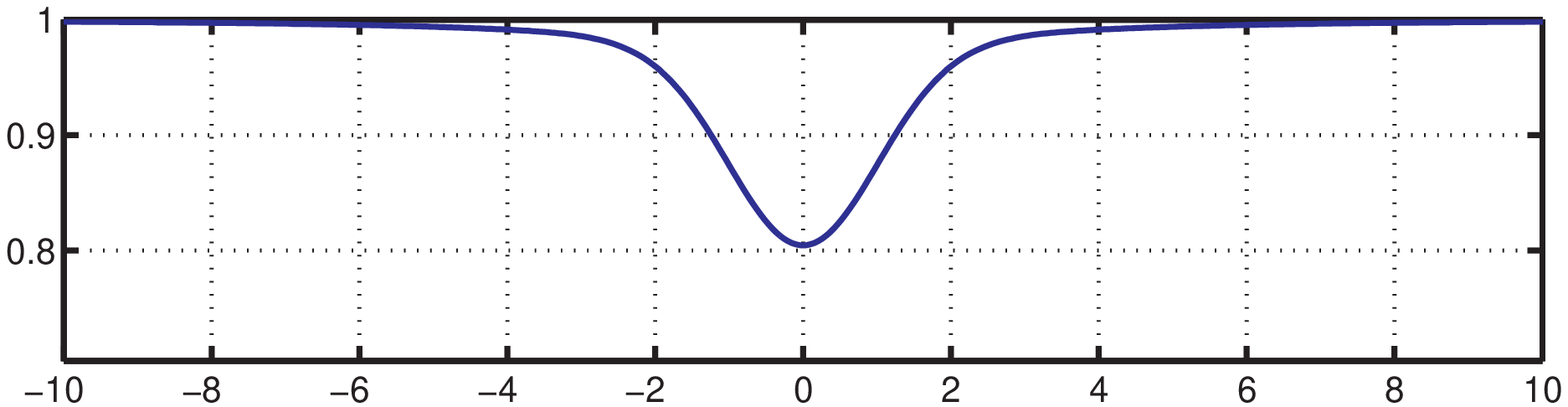}}

\put(-6,12.2){{\Large (a)}} \put(-6,9.4){\Large (b)}
\put(-6,6.6){\Large (c)} \put(-6,3.8){\Large (d)} \put(-6,1){\Large
(e)} \put(-4.3,12.4){ \begin{rotate}{90}\large$F$\end{rotate}}
\put(-4.3,9.6){ \begin{rotate}{90}\large$F$\end{rotate}}
\put(-4.3,6.8){ \begin{rotate}{90}\large$F$\end{rotate}}
\put(-4.3,3.9){ \begin{rotate}{90}\large$F$ \end{rotate}}
\put(-4.3,1.2){ \begin{rotate}{90}\large$F$\end{rotate}}
\put(3.3,-.4){\large$\omega/\omega_0$}
\end{picture}
\end{center}
\caption{Frequency dependance of the Fano function for various
parameter values (a)
$\bar{\kappa}=.05,\;\;\bar{\chi}_{ES}=0,\;\;\bar{\gamma}_L=\bar{\gamma}_R=.5$
and $\lambda=0$, (b)
$\lambda=.4,\;\;\bar{\kappa}=.1,\;\;\bar{\chi}_{ES}=0,\;\;\bar{\gamma}_L=\bar{\gamma}_R=.5$,
(c)
$\lambda=.4,\;\;\bar{\kappa}=.1,\;\;\bar{\chi}_{ES}=3,\;\;\bar{\gamma}_L=.1,\;\;\bar{\gamma}_R=.9$,
(d)
$\lambda=.4,\;\;\bar{\kappa}=3,\;\;\bar{\chi}_{ES}=3,\;\;\bar{\gamma}_L=.1,\;\;\bar{\gamma}_R=.9$,
(e)
$\lambda=.4,\;\;\bar{\kappa}=3,\;\;\bar{\chi}_{ES}=10,\;\;\bar{\gamma}_L=.1,\;\;\bar{\gamma}_R=.9$.
} \label{Fano_spectra}
\end{figure}

\begin{figure}[h!]
\begin{center}
\setlength{\unitlength}{1cm}
\begin{picture}(7,7)
\put(-5,0){\includegraphics[width=70mm]{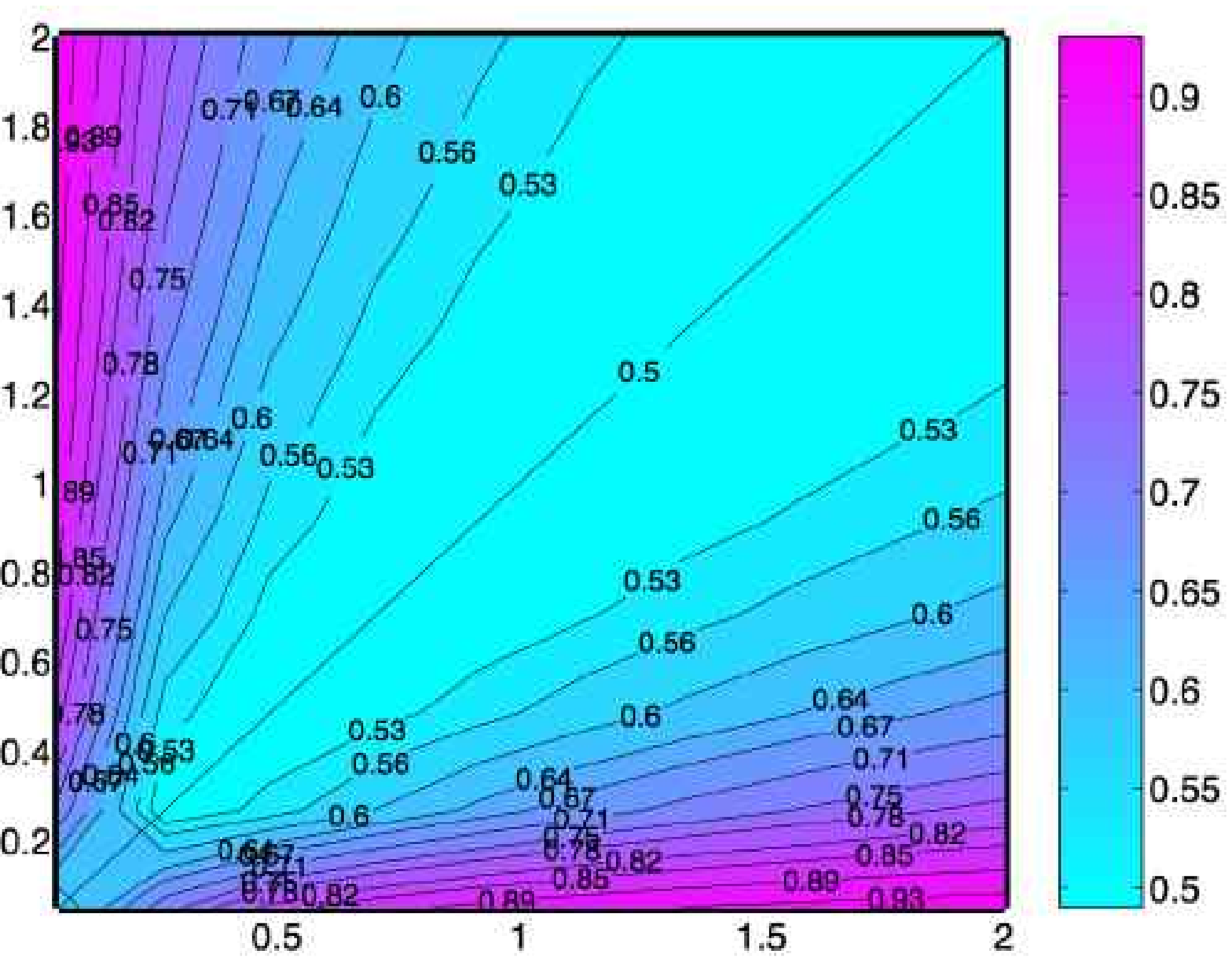}}
\put(4,0){\includegraphics[width=70mm]{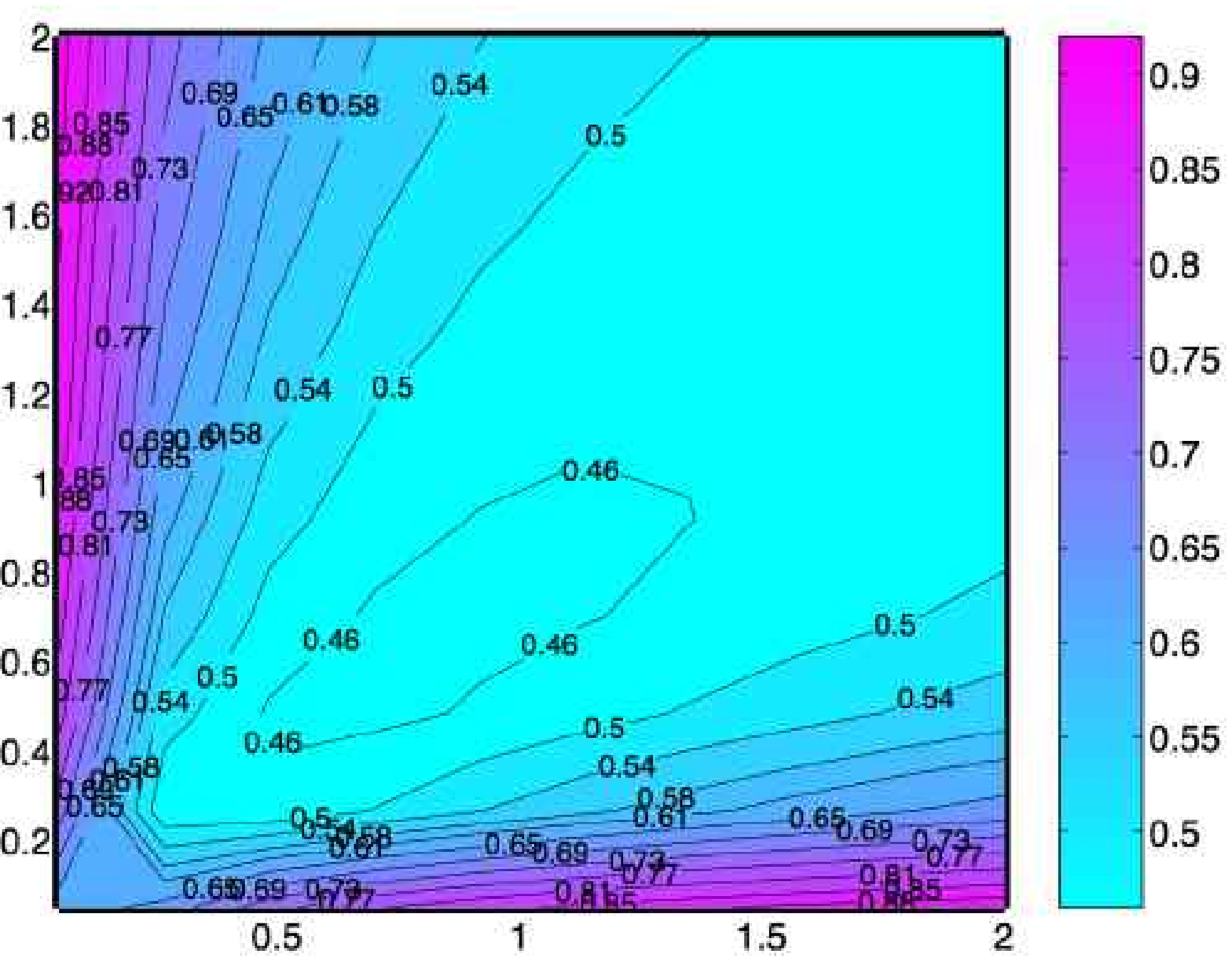}}
\put(-5.2,5.5){\Large (a)} \put(4,5.5){\Large (b)}
\put(-5.6,2.7){\large$\gbr$} \put(-1.7,-.3){\large$\gbl$}
\put(3.1,2.7){\large$\gbr$} \put(7,-.3){\large$\gbl$}
\end{picture}
\end{center}
\caption{Contour plots of the DC Fano factor for the (a) decoupled
system with $\bar{\chi}_{ES}=0,\;\;\bar{\kappa}=.1$, and (b) the
coupled system with
$\lambda=.4,\;\;\bar{\chi}_{ES}=3,\;\;\bar{\kappa}=3$. }
\label{Fano_compare}
\end{figure}

In Figure \ref{Fano_spectra}, we display the current noise spectra
for various frequencies, i.e. \beq
F(\omega/\omega_0)=1-\frac{|S(\omega/\omega_0)|}{2ei^\infty_R}\;\;,
\eeq
 and see that typically the noise is sub-Poissonian.
In Figure \ref{Fano_compare}, we contour plot the DC (or zero
frequency), Fano factors for the decoupled and coupled system and
observe that the coupled system displays stronger anti-bunching when
$\gbl\sim\gbr$, than in the decoupled case. However in the parameter
region where we are interested, $\gbr\gg\gbl$, the DC Fano factors
of the decoupled and coupled system are very similar.

\subsection{Measurement Time}\label{averaging}
\begin{figure}[h!]
\begin{center}
\setlength{\unitlength}{1cm}
\begin{picture}(7,7)
\put(-2,0){\includegraphics[width=120mm]{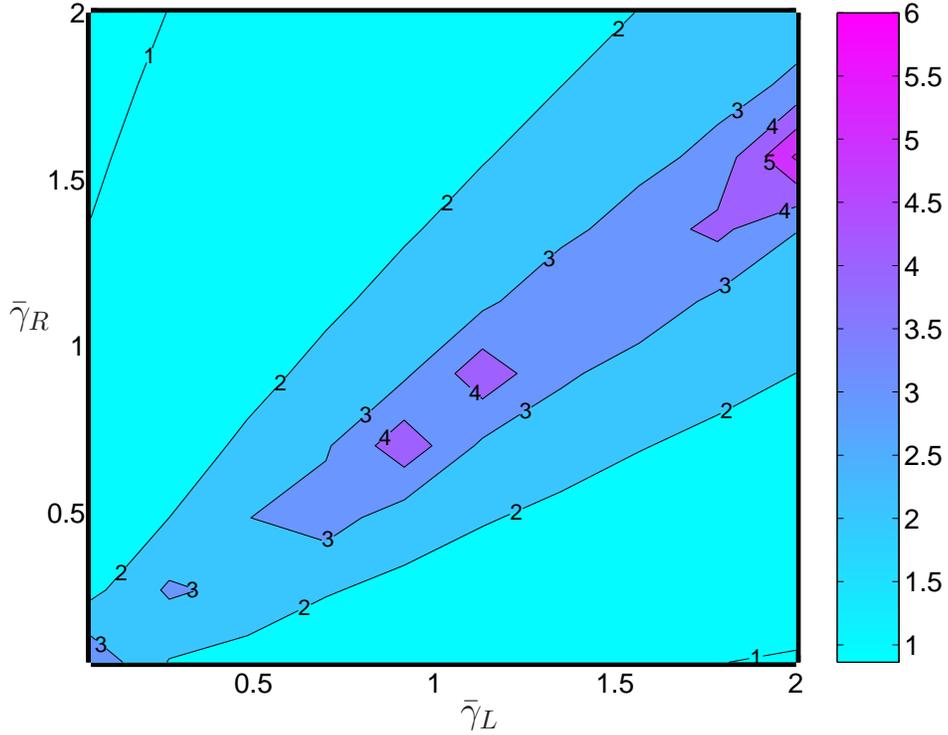}}
\put(3.5,-.3){\Large$\gbl$} \put(-2.5,5){\Large$\gbr$}
\end{picture}
\end{center}
\caption{Contour plot of the $\log_{10}(\tau_m)$, of the measurement
time from Equation (\ref{Korotkoveq}), required to distinguish the
two spin states from the current for the coupled system with
$\lambda=\delta=.4,\;\;\bar{\chi}_{ES}=3,\;\;\bar{\kappa}=3$. }
\label{Korotkov}
\end{figure}

The final quantity we must determine, given the DC current signals
and current noise associated with the two endohedral spin-states, is
the measurement time $\tau_m$, required to differentiate the two
currents $\bar{i}_R^{\uparrow\,\infty}$ and
$\bar{i}_R^{\downarrow\,\infty}$.  This is given by the expression
\cite{Korotkov}, \beq
\tau_m=\frac{(\sqrt{S_1}+\sqrt{S_2})^2}{2(\Delta I)^2}\;\;,
\label{Korotkoveq} \eeq where $S_1$, and $S_2$, are the DC current
noise spectral densities for the two cases of spin up/down while
$\Delta I\equiv |\langle \bar{i}_R^{\uparrow \,
\infty}\rangle-\langle \bar{i}_R^{\downarrow \, \infty}\rangle|$. In
Figure \ref{Korotkov}, we plot $\log_{10} \tau_m$, for various
$\gbl,\;\;\gbr$. We see clearly that when $\gbr\sim \gbl$, $\tau_m$
must be very large to discriminate between the very small difference
in steady-state currents. In the parameter regions we are concerned
with, i.e. $\gbr/\gbl\gg 1$, this measurement time is $\tau_m\sim
10^1-10^2$, in the natural units of the oscillator
($\omega_0=2\pi\times 1$GHz), or a few  tens of nanoseconds. Thus in
this parameter region, the measurement time is very realistic.

\section{Summary}
In summary, we have shown that the right-hand tunnel current of a strongly damped single  electron shuttle operating in the Coulomb blockade regime has very high sensitivity to small equilibrium position displacements of the shuttle. We considered a N@C$_{60}$ molecular shuttle whose endohedral spin can exert a force on the molecule when placed in a large magnetic field gradient. By placing a large nanoscopic magnet nearby, a tiny $\sim\pm  10^{-15}$N force is suffered by the molecular island arising from the endohedral spin state $S_z=\pm 3/2$. This small force alters the
equilibrium position of the island in the presence of the
electrostatic driving force, motional damping, and harmonic
restoring force. Through the extremely sensitive (exponential),
dependence of the conductance on the tunnel separations, the
shift in the equilibrium position of the island profoundly
influences the current flow. We determined the current noise
spectral density and the measurement time  required to distinguish
between the  two DC steady-state current signals (spin up and down),
in the presence of steady-state quantum noise. 

The results we have
found would strongly indicate that there are realistic parameter
regimes where the spin state dependent currents are distinguishable
within several tens of nanoseconds. Thus this device should thus be
capable of spin-detection. Alternatively, similar devices (without an endohedral spin), should be capable of discriminating small, $\sim\pm  10^{-15}$N , static forces on a C$_{60}$ island.

\section{Acknowledgements}
This work was supported via an EC 5FP FET QIPC Project QIPDDF-ROSES.  HSG would like to acknowledge the support from the National Science Council, Taiwan under  Contract No. NSC 94-2112-M-002-028.

\end{document}